\documentclass[%
 reprint,
 amsmath,amssymb,
 aps,
]{revtex4-2}
\usepackage{graphicx}
\usepackage{dcolumn}
\usepackage{bm}

\usepackage{xcolor}
\begin{document}
\preprint{APS/123-QED}

\title{Superconducting Diode Effect \\
“Fundamental Concepts, Material Aspects, and Device Prospects”}

\author{Muhammad Nadeem}
\email{mnadeem@uow.edu.au}
\affiliation{Institute for Superconducting and Electronic Materials (ISEM), Australian Institute for Innovative Materials (AIIM), University of Wollongong, Wollongong, New South Wales 2525, Australia}
\affiliation{ARC Centre of Excellence in Future Low-Energy Electronics Technologies (FLEET), University of Wollongong, Wollongong, New South Wales 2525, Australia}
\author{Michael S. Fuhrer}
\affiliation{School of Physics and Astronomy, Monash University, Clayton, Victoria 3800, Australia.}
\affiliation{ARC Centre of Excellence in Future Low-Energy Electronics Technologies (FLEET), Monash University, Clayton, Victoria 3800, Australia.}
\author{Xiaolin Wang}
\email{xiaolin@uow.edu.au}
\affiliation{Institute for Superconducting and Electronic Materials (ISEM), Australian Institute for Innovative Materials (AIIM), University of Wollongong, Wollongong, New South Wales 2525, Australia.}
\affiliation{ARC Centre of Excellence in Future Low-Energy Electronics Technologies (FLEET), University of Wollongong, Wollongong, New South Wales 2525, Australia.}


\begin{abstract}
Superconducting diode effect, in analogy to the nonreciprocal resistive charge transport in semiconducting diode, is a nonreciprocity of dissipationless supercurrent. Such an exotic phenomenon originates from intertwining between symmetry-constrained supercurrent transport and intrinsic quantum functionalities of helical/chiral superconductors. In this article, research progress of superconducting diode effect including fundamental concepts, material aspects, device prospects, and theoretical/experimental development is reviewed. First, fundamental mechanisms to cause superconducting diode effect including simultaneous space-inversion and time-reversal symmetry breaking, magnetochiral anisotropy, interplay between spin-orbit interaction energy and the characteristic energy scale of supercurrent carriers, and finite-momentum Cooper pairing are discussed. Second, the progress of superconducting diode effect from theoretical predictions to experimental observations are reviewed.
Third, interplay between various system parameters leading to superconducting diode effect with optimal performance is presented. Then, it is explicitly highlighted that nonreciprocity of supercurrent can be characterized either by current-voltage relation obtained from resistive direct-current measurements in the metal-superconductor fluctuation region ($T\approx T_c$) or by current-phase relation and nonreciprocity of superfluid inductance obtained from alternating-current measurements in the superconducting phase ($T<T_c$). Finally, insight into future directions in this active research field is provided with a perspective analysis on intertwining between band-topology and helical superconductivity, which could be useful to steer the engineering of emergent topological superconducting technologies.  
\end{abstract}
\maketitle


Superconducting diode effect (SDE), a recently observed quantum phenomenon in noncentrosymmetric superconductors (SCs) with finite-momentum Cooper pairing, refers to the nonreciprocity of supercurrent \cite{Kun20,Toshiya20,santamaria22}. As depicted by the word `\textit{nonreciprocity}' and `\textit{diode effect}', the system allows supercurrent to flow only in one direction. Similar to the role of semiconducting diode \cite{scaff47,shockley49}, which is one of the central building blocks for (opto-)electronic technologies, e.g., current rectifiers, voltage-controlled oscillators, alternating–direct current converters, LEDs, photodetectors, and solar cells etc., SDE envisions novel device applications in superconducting electronics \cite{tinkham04, braginski19}, superconducting spintronics \cite{linder15,Cai22}, and quantum information and communication technology (QICT) \cite{wendin17,liu19}.\par

After recent observation of SDE for the critical current (fluctuation regime) in symmetric superconductor \cite{Ando20} and for the supercurrent (far below the fluctuation regime) in the Josephson junction (JJ) version \cite{Baumgartner22}, nonreciprocity has emerged as an active research topic in the field of superconductivity. For instance, after seminal observation of SDE in artificially fabricated junction-free superconducting [Nb/V/Ta]$_n$ superlattice, reported first time by F. Ando et al. \cite{Ando20} in 2020, SDE has been experimentally observed in a number of junction-free SCs \cite{Miyasaka21,Narita22,Kenji19,masuko22, Lin22}. Similarly, JJ version of SDE in symmetric Al/InAs-2DEG/Al junction, first reported by Baumgartner et al. \cite{Baumgartner22} in 2022, is followed by SDE experiments on various JJs utilizing different materials acting as a normal barrier or weak link sandwiched between conventional SCs \cite{Jeon22,Golod22,Wu22,Lorenz22,Pal22, DezMrida21}. In addition, observation of SDE has also been demonstrated in engineered superconducting systems, e.g., superconducting thin films with conformal-mapped nanoholes \cite{Lyu22}. The interest in the nonreciprocal supercurrent transport has been further advanced by the recent demonstration of SDE in unconventional/topological superconducting materials. For instance, apart from conventional SCs, SDE has also been observed in unconventional SCs such as magic-angle-twisted bilayer-graphene (MATBLG) \cite{DezMrida21} and small-twist-angle trilayer graphene (STATLG) \cite{Lin22}. Furthermore, SDE has also been demonstrated in topological SCs \cite{Kenji19,masuko22,Pal22} where superconductivity coexists with nontrivial band-topology, e.g., topological JJ \cite{Pal22} where type-II Dirac semimetal NiTe$_2$ is sandwiched between conventional s-wave spin-singlet superconductor Nb, and a topological insulator-superconductor interface such as Bi$_2$Te$_3$/FeTe heterostructure \cite{Kenji19} and Bi$_2$Te$_3$/PdTe$_2$ heterostructure \cite{masuko22}.\par

Following these intriguing experimental observations, and inspired by theoretical work by V. M. Edelstein \cite{Edel89CP,Edel95MCA,Edel96GL}, SDE has been theorized by a number of research groups. For instance, by employing mean-field (MF), Bogoliubov–de Gennes (BdG) and Ginzburg-Landau (GL) theories, theoretical insights have recently been presented for SDE in junction-free bulk SCs \cite{Daido22,Noah22,He22,Ilic22, Scammell22,Zhai22,Zinkl22,Noah21,Legg22,Takasan22} as well as for its JJ version \cite{Hu07,Davydova22,Zhang22-JJ, Baumgartner22-JPCM,Wei22,Pekerten22,Tanaka22d-wave}. Though, the intrinsic mechanism to cause SDE in junction-free SCs is recently clarified, i.e., nonreciprocity of depairing critical current, theoretical modelling for potential spin-orbit coupled bulk SCs is still enjoying its infancy \cite{Daido22,Noah22,He22,Ilic22,Scammell22,Zhai22,Zinkl22,Noah21,Legg22, Takasan22}. In comparison, the underlying mechanism of nonreciprocal supercurrent and SDE is better understood in engineered systems. For instance, diode effect can be engineered in a JJ by controlling Andreev bound states in the normal metal barrier or a weak-link. M. Davydova et al. \cite{Davydova22} showed that such effects in a short JJ can arise from both the Doppler energy shift in the Andreev bound states due to finite-momentum Cooper pairing and the asymmetric current from the continuum of states due to phase-independent contribution. It has also been shown that SDE in JJ \cite{Baumgartner22} and conformal-mapped nanoholes \cite{Lyu22} is well simulated by BdG \cite{Baumgartner22} and time-dependent GL theories \cite{Lyu22}. Furthermore, before even experimental demonstration of SDE in artificial devices \cite{Baumgartner22, Lyu22}, similar nonreciprocal effects have been recognized in several engineered systems \cite{Reynoso08,Zazunov09,Margaris10,Yokoyama14, Campagnano15,Minutillo18,Silaev14,Pal19,Dolcini15,Chen18,Kopasov21}, e.g., conventional JJs \cite{Reynoso08,Zazunov09,Margaris10,Yokoyama14,Campagnano15, Minutillo18}, domain-wall superconducting state \cite{Silaev14}, ferromagnetic JJ with a spin-flipper weak-link acting as a quantized Josephson phase battery \cite{Pal19}, and topological JJs \cite{Dolcini15,Chen18,Kopasov21}. \par

In nonreciprocal quantum materials (NRQM) lacking space-inversion symmetry, direction-selective charge transport can generally be realized whether time-reversal symmetry is broken or not. However, thus far, experimental observation of nonreciprocal supercurrent has only been reported in SCs with simultaneous space-inversion and time-reversal symmetry breaking leading to magnetochiral effects. Space-inversion symmetry is either intrinsically broken or it can be broken by applying an electric field externally. Similarly, time-reversal symmetry can be broken either by applying an external magnetic-field or through intrinsic magnetization, leading to an observation of field-free SDE \cite{Narita22,Jeon22,Golod22,Wu22,Lin22,DezMrida21}. In time-reversal asymmetric SCs, nonreciprocity of supercurrent is guaranteed if the following symmetry-imposed constraint, inducing finite-momentum Cooper pairing, is satisfied: both the orientation along which inversion/mirror symmetry is broken and the direction along which (super)current is flowing must be perpendicular to the magnetic-field orientation or magnetization polarization. Thus, owning to the magnetochiral effects, nonreciprocal supercurrent can be switched by reversing the orientation/polarization of magnetic-field/magnetization.\par

\begin{figure*}
\includegraphics[scale=0.41]{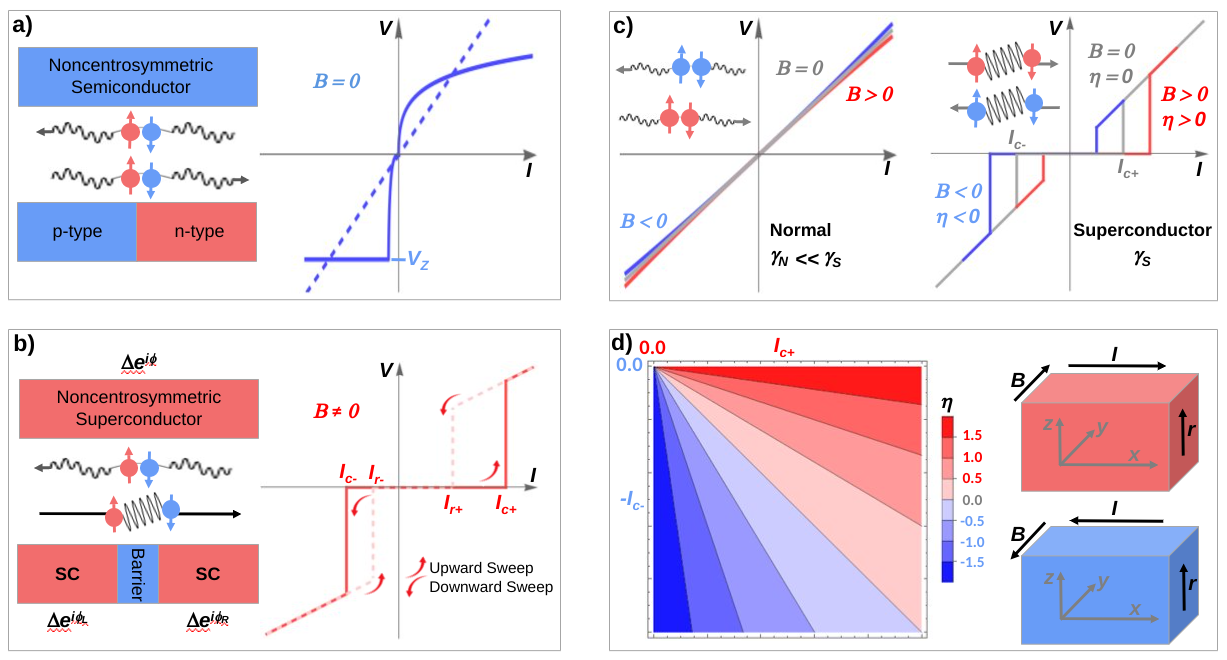}
\caption{\label{SDE}\textbf{Diode effect in semiconductors and SCs.} Here straight black lines represent supercurrent flowing due to coherent Cooper pairs while the wiggly black lines represent normal current flowing due to depaired electrons. \textbf{(a)} Diode effects in noncentrosymmetric bulk semiconductors and pn junctions. (\textit{Left}) Diode effects, such as rectification, can be realized in a junction-free and noncentrosymmetric bulk electrical conductor (top) and at a semiconducting pn junction (bottom). (\textit{Right}) I-V curve for noncentrosymmetric bulk semiconductors (dashed) and pn junctions (solid). \textbf{(b)} Superconducting diode effect in junction-free noncentrosymmetric bulk crystals and JJs. (\textit{Left}) SDE in a bulk crystal with an order parameter $\Delta e^{i\phi}$ (top) and SDE in a JJ between two SCs with phases $\phi_L$ and $\phi_R$, which are separated by a normal barrier (bottom). (\textit{Right}) I-V curves for SDE in a bulk crystal (solid red lines) and a JJ (solid red lines and dashed red curves) show that a SDE occurs when $I_{c+}\ne I_{c-}$ while the superconductor becomes a normal metal when $I$ is larger than the critical current $I_{c+}$ (along positive direction) or $I_{c-}$ (along negative direction). For a JJ version, $I_{r+}$ and $I_{r-}$ represent two critical return currents in the downward sweep measurements and leads to another non-reciprocal effect when $I_{r+}\ne I_{r-}$. \textbf{(c)} Schematic illustration of nonreciprocal current/supercurrent in noncentrosymmetric bulk crystals. (\textit{Left}) In the normal state of a noncentrosymmetric crystal, whose MCA coefficient ($\gamma_N$) is usually tiny, non-linear I-V curves (red and blue) show a small deviation form the linear I-V curve (gray), indicating a small nonreciprocal current. (\textit{Right}) In the fluctuation regime (resistive superconducting state) of a noncentrosymmetric crystal, whose MCA coefficient ($\gamma_S$) becomes much larger than that of the normal state, I-V curve shows large nonreciprocal current below the critical current ($I_c$), whereas it resembles to that for the normal state and remains unchanged at $I>I_c$. Here $\eta$ represents efficiency of a superconducting diode which changes its sign when the polarity of magnetic field (B) is reversed. \textbf{(d)} Strength of superconducting diode effect. (\textit{Left}) The superconducting rectification becomes maximal when $+I_c$ (or $-I_c$) remains finite but $-I_c$ (or $+I_c$) becomes zero. As defined by the diode efficiency, equation (\ref{Eff}), the maximum difference of critical depairing currents $+I_c$ and $-I_c$ can be about a factor of 2. (\textit{Right}) In the junction-free noncentrosymmetric bulk material, rectification can be induced by applying magnetic field perpendicular to the directions of both the polar axis and the current.}
\end{figure*}

In this article, recent theoretical and experimental progress on SDE including fundamental concepts, material aspects, and device prospects, is reviewed. In section \ref{Mech}, fundamental concepts and various mechanisms to cause SDE, i.e., nonreciprocal charge transport, magnetochiral anisotropy (MCA), breaking of space-inversion/time-reversal symmetry, type of associated spin-orbit interaction (SOI), origin and orientation of magnetization, upper critical field, and finite-momentum Cooper pairing or helical/chiral superconductivity are discussed. We highlighted how all these mechanisms are closely related with each other and, especially, their intertwining with SOI, which is a fundamental relativistic quantum functionality. In section \ref{Theory}, theoretical progress is reviewed for bulk and engineered SCs. Section \ref{Materials} covers the material aspects. Various superconducting materials, in which observation of SDE has been reported, are classified based on the geometric structure of diode device, nature of SOI, origin and orientation of magnetization, and their topological character. Section \ref{Efficiency} demonstrates how the strength/efficiency of SDE depend on a range of parameters such as magnetic field, temperature, Cooper pairing momentum, SOI, chemical potential \cite{He22}, next-nearest neighbour hopping \cite{Daido22}, disorder \cite{Ilic22}, and design or characteristics of a JJ \cite{Baumgartner22,Tanaka22d-wave}. Section \ref{Obser} covers two main techniques employed for the observation of SDE: nonreciprocity of critical current via resistive direct-current (dc) measurements and nonreciprocity of supercurrent via inductive alternating-current (ac) measurements. In section \ref{outlook}, the article is concluded with a perspective on future directions and device prospects of SDE. Since SDE is a novel quantum mechanical phenomenon and the system hosting this effect may prove to be a key component of emergent quantum technologies, we hope this review article may shed light on profound understanding of fundamental mechanism/concepts of SDE and may allow search of novel superconducting systems for emergent superconducting technologies such as electronics, spintronics, optoelectronics, and fault-tolerant quantum computing. \par

\section{\label{Mech}Mechanisms of superconducting diode effect}
The origin of SDE manifests in a number of physical phenomena, imposed by transport mechanisms, symmetry constraints, and underlying quantum functionalities of superconducting materials. In this section, it is explicitly demonstrated how nonreciprocity of supercurrent is intertwined with underlying symmetries of noncentrosymmetric systems, e.g., nonreciprocity driven by MCA in time-reversal asymmetric systems and that induced by shift current or Coulomb interactions in time-reversal symmetric systems. It is also highlighted how nonreciprocity of supercurrent is associated with nonreciprocal behaviour of physical quantities characterizing current-voltage (I-V) and current-phase relation (CPR), e.g., resistance and inductance respectively. Quantum functionalities of SCs, such as SOI, Berry phase, band-topology and their effects on the SDE efficiency are also discussed. Finally, intertwining between nonreciprocity and helical superconductivity with finite-momentum spin-singlet or spin-triplet Cooping pairing is presented. \par

\subsection{Nonreciprocity and magnetochirality}
In condensed matters, nonreciprocity refers to the spatial-dependence of physical quantities. A prototypical example of nonreciprocal transport is a diode effect which refers to a highly direction-selective electron transport in systems with a lack of spatial inversion center. Until recently, nonracirocity was thought to be a transport phenomenon associated with dissipative materials. For instance, in conventional semiconductors, where resistance is the nonreciprocal quantity, nonreciprocity refers to charge transport that is sensitive to the polarity of current or bias potential. Such nonreciprocal charge transport leads to a diode effect in a spatially asymmetric pn junction \cite{scaff47,shockley49}, in which, spatial asymmetry of the junction is associated with electron-hole asymmetry across the contact of n- and p-type semiconductors.\par

In modern quantum condensed matter physics, in addition to electron-hole asymmetric junctions, nonreciprocal charge transport can be induced in spatially symmetric devices, in which resistance is direction-selective when inversion and/or time-reversal symmetry are broken. This can be realized, for instance, by externally applying an electric and/or a magnetic field orthogonal to each other and to the direction along which current is traversing. It implies that nonreciprocal transport can be treated as a bulk property of noncentrosymmetric quantum materials \cite{Tokura18,Toshiya21}. In noncentrosymmetric systems, i.e., in which inversion symmetry is broken, nonreciprocal responses can be classified into four categories \cite{Tokura18}: (i) linear- and (ii) nonlinear-response in time-reversal symmetric systems, and (iii) linear-  and (iv) nonlinear-response in time-reversal asymmetric systems.\par

When both inversion and time-reversal symmetry are simultaneously broken, a closely related phenomenon leading to nonlinear nonreciprocal response is MCA \cite{Edel95MCA,Edel96GL,Rikken01,Krstic02,Pop14,Rikken05,Morimoto16, Ideue17,Wakatsuki18,Hoshino18,Wakatsuki17,Qin17,Kenji19,Choe19,Yuki20}. In the linear response regime of noncentrosymmetric systems, broken time-reversal symmetry produces finite magnetochiral effect, as recognized by the Onsager’s reciprocal theorem \cite{onsager31,kubo57,Tokura18,Hoshino18}, and the longitudinal transport coefficients become dependent on the polarity of the current. The Onsager’s reciprocal theorem, and thus the magnetochiral effect and direction-selective transport, can be generalized to the nonlinear regime of both (semi)conductors \cite{Rikken01,Rikken05} and SCs \cite{Edel95MCA,Edel96GL}.

\subsubsection{Nonreciprocity of supercurrent}
In 1996, before even prediction/observation of nonreciprocity in (semi)conductors by Rikken et al. \cite{Rikken01,Rikken05}, V. M. Edelstein \cite{Edel96GL} proposed nonreciprocity in the critical supercurrent. Followed by his earlier work characterizing Cooper pairing in noncentrosymmetric SCs \cite{Edel89CP} and describing magnetoelectric effects in polar SCs \cite{Edel95MCA}, V. M. Edelstein \cite{Edel96GL} proposed that if the mixed product $(\boldsymbol{c}\times \boldsymbol{B})\cdot \boldsymbol{\hat j_c}$ is non-vanishing in polar SCs, then the magnitude of the critical current $j_c(B)$ depends on the sign of this mixed product, i.e., the critical current appears to be different for two opposite directions. By employing GL theory for a thin film of polar superconductor, expression for the nonreciprocity in the critical current reads \cite{Edel96GL}
\begin{equation}
j_c(B)=j_c(0)[1+\gamma_j(\boldsymbol{c}\times \boldsymbol{B})\cdot \boldsymbol{\hat j}]\;\label{J}
\end{equation}
Here $c$ is the unit vector along the polar axis, $\boldsymbol{\hat j}$ is the unit vector along the supercurrent, and $B$ is an in-plane magnetic field. The exact expression for the observable can be found in the reference \cite{Edel96GL}.

MCA and nonreciprocity has been observed in (semi)conductors \cite{Krstic02,Pop14,Rikken05,Morimoto16,Ideue17,Choe19} that show resistive current as well as in SCs \cite{Wakatsuki17,Qin17,Kenji19,Yuki20} that display dissipationless supercurrent. So a question arises naturally: how nonreciprocity can uniquely be defined in these two systems with completely contrasting behaviour? As first pointed by Rikken et al. \cite{Rikken01}, when both inversion and time-reversal symmetries are broken, the finite MCA coefficient $\gamma$ gives rise to different resistance for electric currents traversing in different (opposite) directions. That is, MCA can be defined as the inequivalence of $R(+I)$ and $R(-I)$. In (semi)conductors, resistances along opposite directions differ, i.e. $R(+I)\ne R(-I)$, but both $R(+I)$ and $R(-I)$ normally take finite values. On the other hand, in SCs, such a situation becomes more drastic: either one of $R(\pm I)$ remains finite while the other vanishes completely. \par

With this consideration in SCs, it becomes more appropriate to define nonreciprocity in terms of (super)current. That is, as shown in figure [\ref{SDE}], nonreciprocity in SCs means supercurrent flows along one direction while a normal current along the other(opposite). Observation of such a situation is more probable near critical temperature $T_c$, i,e., in the fluctuation regime of metal-superconductor resistive transition, where the critical current is different along opposite directions, i.e. $I_{c+}\ne I_{c-}$. Thus, if the current is tuned between $I_{c+}$ and $I_{c-}$, the system displays zero resistance for supercurrent but nonzero for the normal current. \par

It can be understood how conductance varies while going from normal to a superconducting phase. The linear resistance $R_0$ is normally scaled by the Fermi energy $E_F$, i.e., the kinetic energy of the electrons, while the MCA coefficient $\gamma$ depends upon the strength of SOI and the magnetic field. Correspondingly, nonlinear resistance induced by MCA may be treated as a perturbation to $R_0$. In the normal conducting phase, because the SOI energy ($E_{soi}$) and the Zeeman energy ($\mu_B B$) is usually much smaller (by many orders of magnitude) than $E_F$, MCA coefficient $\gamma\rightarrow\gamma_N$ is typically very tiny, usually of the order of $\sim10^{-3}$ to $10^{-2}$ T$^{-1}$ A$^{-1}$ in typical metals \cite{Rikken01,Pop14,Wakatsuki17}. However, as the superconducting phase develops, superconducting transition temperature $T_c$ or the superconducting gap $\Delta_{sc}$ appear as a new energy scale. That is, the energy scale in SCs, to which the strength of SOI has to be compared with, is superconducting gap and not the Fermi energy. Since the energy scale ($\sim$meV) in the SCs is much smaller than the Fermi energy ($\sim$eV) in metals, the effects of SOI and Zeeman energy greatly enhance in the superconducting phase \cite{Wakatsuki17,Hoshino18}. As a result, near the superconducting transition temperature $T\gtrsim T_c$, MCA coefficient becomes reasonably large \cite{Tokura18} and, thus, the paraconductivity \cite{Schmid69} above $T_c$  becomes nonreciprocal. In the superconducting fluctuation region, i.e. when $T\rightarrow T_c$ and the superconducting order parameter $\Delta_{sc}$ develops, a sizable enhancement in MCA coefficient $\gamma_S$ is found (ref.\cite{Tokura18,Wakatsuki17,Hoshino18,Kenji19}) and a robust non-reciprocal charge transport is demonstrated in noncentrosymmetric SCs \cite{Wakatsuki17,Yuki20}. For instance, by employing GL theory for an Ising type SC MoS$_2$, R. Wakatsuki et al. \cite{Wakatsuki17} showed that the ratio of MCA coefficients in the superconducting resistive region ($\gamma_S$) and the normal resistive region ($\gamma_N$) is quite large
\begin{equation}
\frac{\gamma_S}{\gamma_N}\sim \left(\frac{E_F}{k_BT_c}\right)^3
\end{equation}
Such anomalous enhancement of the MCA coefficient, as it is associated with the energy scale difference between the superconducting gap and the Fermi energy, can be considered an intrinsic feature of both Rashba and Ising type noncentrosymmetric SCs \cite{Hoshino18}. However, mainly due to a gradual decrease in the linear resistance $R_0$ during the metal-superconducting transition, $R_0$ remains larger (by orders of magnitude) than the nonlinear resistance in low-dimensional superconducting materials such as MoS$_2$ (ref. \cite{Wakatsuki17}), WS$_2$ (ref. \cite{Qin17}) and Bi$_2$Te$_3$/FeTe (ref. \cite{Kenji19}). As a result, low rectification ratio in these superconducting materials does not suffice for device implementation. In this regard, it is highly desired to search for novel mechanisms/principles to enlarge the rectification effect and guide the design of efficient SDE.

\subsubsection{From resistance to supercurrent}
Rikken et al. \cite{Rikken01,Rikken05} generalized the Onsager’s reciprocal theorem to the nonlinear regime and gave a heuristic argument for nonreciprocity and MCA in two-dimensional diffusive conductors. In their seminal proposal of MCA in (semi)conductors, Rikken et al. \cite{Rikken01} suggested that nonreciprocal nonlinear resistive response, characterized by the directional IV-characteristics, can be described by a current-dependent resistance $R(I)$ as
\begin{equation}
R(I)=R_0[1+\beta B^2+\gamma(\boldsymbol{B}\times \boldsymbol{r}) \cdot \boldsymbol{I}] \;\label{R}
\end{equation}
Here R, B, and I are the resistance, magnetic field, and the electric current, respectively. The unit vector $\boldsymbol{r}$ represents the direction along which mirror symmetry is broken. On the right-hand side, first term is the resistance at zero magnetic field, second term denotes the normal magnetoresistance, and the third term corresponds to the MCA. Dependence of MCA coefficient $\gamma$ on electric current, magnetic field, as well as their mutual orientation, relative to the direction along which mirror symmetry is broken, allows us to access various functionalities and aspects of noncentrosymmetric materials. \par

First, dependence of MCA on electric current leads to a current-dependent resistance which generally causes a nonlinear nonreciprocal transport, i.e. nonlinear voltage-drop. Such nonlinear nonreciprocal transport can be detected by measuring the second harmonic signal through lock-in techniques, see further details in section (\ref{MCA-R}). Second, dependence of MCA on magnetic field implies that its coefficient $\gamma$ remains non-zero only when time-reversal symmetry is broken. In addition, the orientation of magnetic field must be orthogonal to both current and the direction along which mirror symmetry is broken. It implies that, not only finite magnetic field is required, but its orientation is also important depending upon the nature of SOI associated broken mirror symmetry. Here we discuss the key mechanisms associated with nonreciprocity in superconducting systems. 

The conventional semiconducting diode is not favorable for energy-efficient technologies with ultralow power consumption. At high temperatures relevant for thermionic transport, owning to their finite resistance, energy loss is inevitable in semiconductors. At low (sub-Kelvin) temperatures, on the other hand, relevant for cryogenic electronics \cite{braginski19} and ultrasensitive (sub-THz frequencies) optoelectronics and detection \cite{farrah19}, semiconductors cease to work due to their large energy gap. Therefore, owning to their dissipationless supercurrents, intrinsically low impedance and thereby very high rectification of supercurrents, and low energy scales associated with superconducting gap ($\sim$meV) as compared to semiconductor energy gap ($\sim$eV), a superconducting diode is highly desired for energy-efficient cryogenic electronic/optoelectronic devices \cite{tinkham04,braginski19}. However, as broken electron-hole symmetry is required, physical realization of a junction-free superconducting diode turns out to be difficult with electron-hole symmetric Bardeen–Cooper–Schrieffer (BCS) superconducting state. \par

In light of this, SCs with broken spatial-inversion and time-reversal symmetry can offer bright perspectives for supercurrent diode effect via MCA. However, for the implementation of a simplest possible device displaying SDE intrinsically, it is worthy to pin down intertwining between superconductivity and MCA. First, unlike rectification due to self-field effects in asymmetric superconducting quantum interference devices (SQUIDs) \cite{fulton72,barone82}, MCA is expected to induce an intrinsic SDE in symmetric devices with spatially homogeneous supercurrent density. Second, intertwining between superconductivity and MCA can lead to a spin-filtering diode effect in a spin-selective Al/EuS/Cu superconducting tunnel junction \cite{Strambini22} and thus superconducting spintronic technologies \cite{linder15}. However, even such promising ferromagnetic superconducting structure, in which electron-hole symmetry can possibly be broken when both spin-filtering and spin-splitting are present to induce opposite shift in BCS density of state (DOS), are not desired for the intrinsic SDE with nonreciprocal supercurrent transport. Finally, we could see the light at the end of the tunnel: Intrinsic SDE with nonreciprocal supercurrent transport can be realized in a helical superconductor with finite-momentum Cooper pairing which can be induced by antisymmetric Rashba/Ising SOI and Zeeman exchange spin-splitting. Further details on this key mechanism are discussed in section \ref{Helical}.\par

\subsubsection{From inductance to supercurrent}
Nonreciprocity in the fluctuation regime of metal-superconductor resistive transition confines SDE to a narrow temperature window near $T_c$. Baumgartner et al. \cite{Baumgartner22} pointed that the temperature window in which MCA coefficient becomes sizeable must be widened for a sustainable fabrication of devices showing SDE. To achieve this milestone, the authors demonstrated supercurrent rectification in the superconducting phase, i.e., far below the transition temperature $T_c$. Since d.c. measurement of resistance–current (R–I) curve is not viable at low temperatures, as the resistance vanishes, supercurrent response to an alternating-current (a.c.) excitation is studied, which is described by its superfluid stiffness, and thus, can be detected through kinetic inductance measurements.\par
 
If mirror symmetry is broken along out-of-plane direction ($\hat{e}_z$), whereas the current I and magnetic field B are directed in-plane, MCA or nonreciprocity for the superfluid can be described by an equation similar to that for the resistance (\ref{R}), i.e.,
\begin{equation}
L(I)=L_0[1+\gamma_L \hat{e}_z (\boldsymbol{B} \times \boldsymbol{I})] \;\label{L}
\end{equation}
Here resistance ($R$) is substituted for the kinetic inductance ($L$). The nonraciproocity in supercurrent could then be characterized by a new observable, i.e., MCA coefficient $\gamma_L$.

\subsection{Nonreciprocity without magnetochirality}
In noncentrosymmetric but time-reversal symmetric systems, nonreciprocal nonlinear response can be realized via shift current (photovoltaic effect) \cite{von81,sipe00,Nagaosa17-Rev}, via Coulomb interactions \cite{morimoto18}, and asymmetric Hall effect of vortices and antivortices \cite{itahashi22}. Shift current is a nontrivial contribution by the Berry phase of the electronic states \cite{Xiao-RMP}. That is, unlike conventional charge transport which comes form intraband transition \cite{von81,sipe00,Nagaosa17-Rev} and depends only on the energy dispersion, interband shift current depends not only on the energy dispersion but also on the Bloch wavefunction and plays an essential role in modern quantum transport phenomena \cite{Xiao-RMP,nagaosa10-RMP}. Followed by theoretical proposals \cite{von81,sipe00}, shift current has been studied for semiconductor (GaAs) \cite{cote02}, ferroelectric semiconductor (SbSI) \cite{ogawa17}, and Dirac surface states of a 3D topological insulator (Bi$_2$X$_3$(X=Te, Se)) with a hexagonal warping \cite{kim17}. It shows that shift current is an ubiquitous phenomenon in noncentrosymmetric quantum materials, and the nonreciprocal nonlinear response can also be realized without breaking of time-reversal symmetry. 

T. Morimoto and N. Nagaosa \cite{morimoto18} theoretically showed that nonreciprocal nonlinear I–V characteristics can be induced by electron correlations in noncentrosymmetric multiband systems without time-reversal symmetry breaking. According to general symmetry considerations, nonreciprocal nonlinear response in such time-reversal symmetric systems is generally constrained by the presence of two ingredients: (i) dissipation, and (ii) interactions (e.g., electron-electron and electron-phonon interactions). First, generalization of Onsager’s reciprocal theorem to nonlinear current responses shows that dissipation is crucial for nonreciprocity. Second, gauge invariant formulation of Keldysh Green’s function shows that nonreciprocity disappears without interactions. A general formula of the nonreciprocity ratio ($\gamma_c$), and derived by employing nonequilibrium Green’s functions for two-band systems with onsite Coulomb interaction, reads \cite{morimoto18}
\begin{equation}
\gamma_c=\frac{\delta J}{J}\simeq \frac{U}{E_{g,k_F}}\frac{eEa}{W}
\end{equation}
where $U$ is Coulomb interaction energy ($\gamma_c\rightarrow0$ for $U\rightarrow0$), $E_{g,k_F}$ is the band gap, $k_F$ is the Fermi momentum, $e$ is charge of electron, $E$ is the applied electric field, $a$ is the lattice constant, and $W$ is the bandwidth. Here $J$ is the linear current response (the part of current response proportional to $E$) while $\delta J$ is the nonlinear current response (the part of current response proportional to $E^2$). When $U \approx E_{g,k_F}$, nonreciprocal response can be estimated by quantifying the ratio $eEa/W$ between the electric potential ($eEa$) in the unit cell and the bandwidth ($W$).

First the nonreciprocity induced by electron correlation \cite{morimoto18} is relatively smaller than that induced by MCA, in both typical metals \cite{Rikken01,Pop14} as well as resistive semiconductors \cite{Rikken05}. Second, the requirement of dissipation means nonreciprocal response induced by Coulomb interactions is only measurable in the resistive fluctuation regime of metal-superconductor transition, and not in the superconducting phase below transition temperature. On the other hand, nonreciprocity of supercurrent by asymmetric Hall effect of vortices and antivortices in time-reversal symmetric trigonal superconductors (PbTaSe$_2$) \cite{itahashi22} promise another nonlinear transport phenomena to study SDE. However, thus far, experimental observation of nonreciprocity of supercurrent has only been reported in noncentrosymmetric systems with broken time-reversal symmetry, while the observation of supercurrent nonreciprocity in time-reversal symmetric SCs is scarce.

\subsection{Role of spin-orbit coupling}
Apart from the strength of SOI, since broken inversion symmetry is assumed/required ($\gamma=0$ for centrosymmetric systems), MCA coefficient $\gamma$ also depends on the nature of associated SOI. That is, based on the lattice symmetry, finite $\gamma$ may be realized in noncentrosymmetric condensed matter systems \cite{Toshiya21} such as polar or Rashba SCs and trigonal or Ising SCs. In polar systems, where Rashba SOI generated from broken $\mathcal{M}_z$ and electron's spin is locked to in-plane orientations, nonreciprocal supercurrent is controlled by an in-plane magnetic field. On the other hand, in trigonal systems with $D_{3h}$ symmetry, where Ising or valley-Zeeman SOI is originated from broken $\mathcal{M}_{x/y}$ and electron's spin is locked to out-of-plane orientations, nonreciprocal supercurrent is controlled by an out-of-plane magnetic field. \par

In addition, it would be interesting to study effects on SDE due to a crossover between various SOI types associated with broken inversion symmetry. For instance, Baumgartner et al. \cite{Baumgartner22-JPCM} studied effects of Rashba and Dresselhaus SOI on supercurrent rectification and MCA by fabricating Al/InAs-2DEG/Al ballistic JJs. Similarly, Pekerten et al. \cite{Pekerten22} studied an interplay between Rashba and Dresselhaus SOI and investigated effects of magnetic and crystalline anisotropies on the topological superconductivity in JJs. If only Rashba-type SOI is present in the JJs, the topological phase diagram strongly depends on the magnetic field orientation but remains insensitive to the supercurrent polarity. On the other hand, when both Rashba- and Dresselhaus-type SOIs coexist, the phase diagram exhibits a strong dependence on the magnetic field as well as junction crystallographic orientations. These studies illustrate the role of SOI, both for the material search leading to SDE with the best performance and probing phase diagram of topological/helical SCs. \par

Furthermore, H. Yi recently showed a crossover from Ising- to Rashba-type superconductivity in epitaxial topological insulator and monolayer Ising superconductor heterostructure \cite{yi22crossover} (Bi$_2$Se$_3$/NbSe$_2$). By altering the thickness of Bi$_2$Se$_3$ film, emergence of topological superconductivity coincides with a considerable suppression of the upper critical in-plane magnetic field. While the former transition is marked by the emergence of spin-non-degenerate surface states and Rashba-type quantum-well bands in the bulk, the later signatures a crossover from Ising- to Rashba-type superconductivity. This system represents a classic example and sheds light on the role of SOI while searching new systems to engineer SDE.\par 

Based on the above discussion, one can conclude that Ising/trigonal topological SCs, such as NbSe$_2$ which display exceptional upper critical-fields exceeding the Pauli limit \cite{Xi16,Xing17,De18}, can be identified as suitable materials for the realization of SDE via magnetic field driven MCA. On the other hand, owning to the nontrivial Berry phase intertwined with band topology, time-reversal symmetric polar/Rashba SCs can be identified as promising materials for the realization of SDE via shift current. This qualitative analogy needs further quantitative investigation, as the performance of SDE also depends upon the strength of SOI, interband transition, and photoresponse etc.

\subsection{\label{Helical}Helical superconductivity}
To observe SDE via MCA in noncentrosymmetric superconductor, breaking of time-reversal reversal symmetry ($\mathcal{T}$) is necessary but not sufficient. First, SDE is not necessarily present in all the magnetic SCs but rather the orientation of magnetic field or magnetization should be such that it breaks all possible inversion symmetries $\mathcal{P}_i$ ($i=x,y,z$). Second, time-reversal reversal symmetry should be broken such that a finite-momentum Cooper pairing or a helical superconductivity emerges. Third, magnetic field (or magnetization) should have a component perpendicular to the polarity of applied current such that finite pairing momentum emerges parallel/anti-parallel to the current direction. In this section, after a brief overview of helical superconductivity, desired orientation of magnetic field or magnetization, and its intertwining with the nature of SOI, polarity of applied current, direction along which structural mirror symmetry is broken, and the momentum space orientation of Cooper pairing momentum is discussed. \par

\begin{figure*}
\includegraphics[scale=0.48]{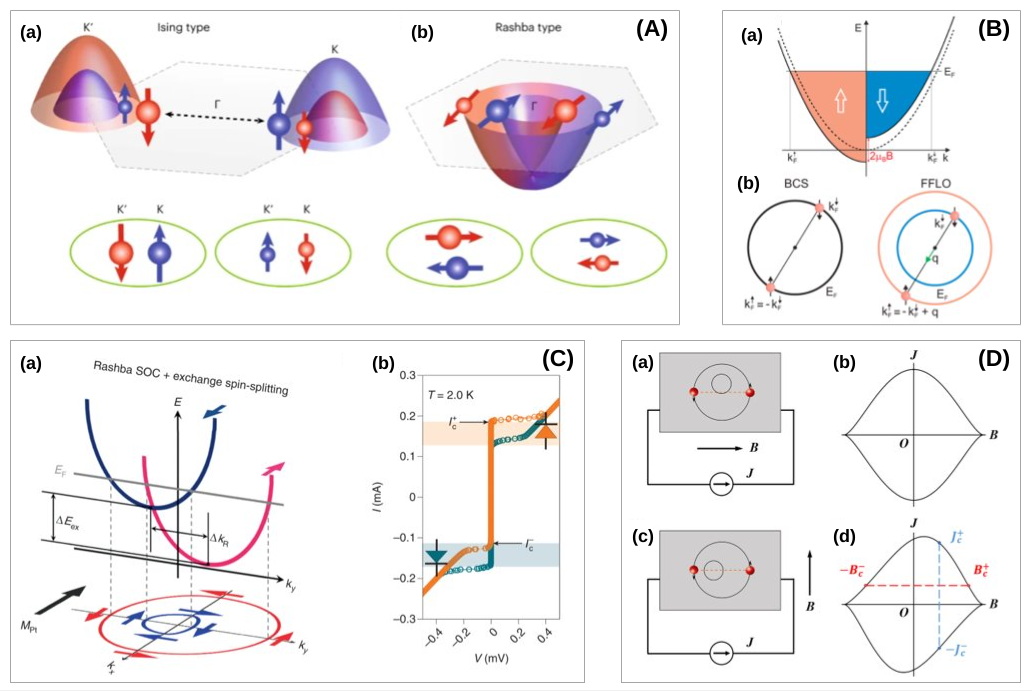}
\caption{\label{HSC}\textbf{abc} \textbf{(A)} Schematics of the Ising- and Rashba-type superconducting pairing symmetry. \textbf{a} Ising-type pairing symmetry originates from spin-singlet Cooper pairs formed between the electrons near the K and K$^\prime$ valleys with opposite spins pinned to the out-of-plane direction. \textbf{(b)} Rashba-type pairing symmetry originates from spin-singlet Cooper pairs formed between the electrons near the $\Gamma$ point with opposite momentum and opposite spins pinned to the in-plane direction. Figure (A) is reproduced with permission from ref. \cite{yi22crossover}. \textbf{(B)} (a) Schematic sketch showing magnetic field driven spin-splitting of free-electron parabola, inducing Pauli paramagnetism, and leading to different Fermi momenta for spin-up ($k^\uparrow_F$) and spin-down ($k^\downarrow_F$) electrons. (b) Schematic representation of the conventional spin-singlet BCS pairing state (left) with zero center-of-mass momentum and the spin-singlet FFLO pairing state (right) with a finite center-of-mass momentum (q). The red (blue) circle represents the Fermi surface for electrons with spin-up (spin-down). Figure (B) is reproduced with permission from ref. \cite{Wosnitza18Rev}. \textbf{(C)} Band splitting and Fermi contours under Rashba SOI and exchange field in a JJ Nb/Pt/Nb with a Pt barrier proximity-magnetized by a ferrimagnetic insulating Y$_3$Fe$_5$O$_{12}$ (YIG) film. \textbf{a} Rashab SOI splits the conduction bands laterally (along momentum (k) axis) by $\Delta k_R$ while the Zeeman exchange field splits them vertically (along energy (E) axis) by $\Delta E_{ex}$ such that the Kramers degeneracy is removed. Here $E_F$ represents the Fermi level while $k_{x/y}$ stands for in-plane momentum components. \textbf{b}I-V curve representing SDE at T = 2 K ($<T_c$) for different orientations of the Pt magnetization (M$_{Pt}$), parallel (yellow) and antiparallel (cyan) with respect to the x-axis. Here Pt magnetization orientations, and, thus the the direction of the exchange field, reverses when the magnetization orientation of the proximity-coupled YIG is inverted. The diode symbols in the yellow (cyan) shaded regime indicates that the Josephson supercurrent flows only in the positive (negative) y-direction. Figure (C) is reproduced with permission from ref. \cite{Jeon22}, Springer Nature Ltd. \textbf{(D)} Supercurrent diode effect under external current source J and in-plane magnetic field B in a noncentrosymmetric Rashba superconductor. (a and c) Schematics of device plots showing Rashba- and Zeeman-split normal state Fermi surfaces (denoted by circles) and the directions along which J and B are applied. (b and d) Schematic phase diagrams in the B-J plane corresponding to device configurations shown in (A and B), respectively. Figure (D) is taken from ref. \cite{Noah22}.}
\end{figure*}

\subsubsection{Fulde–Ferrell–Larkin–Ovchinnikov state}
In the field of conventional superconductivity, following from the fact that Cooper pairing is formed between Kramers partners and most known conventional SCs are characterized by the Bardeen–Cooper–Schrieffer theory \cite{BCS57}, presence of time-reversal symmetry is a key ingredient and the preserved Kramers degeneracy is the fundamental reason/criterion that stabilize superconducting phase in so many systems at sufficiently low temperatures  \cite{ginzburg57fm,matthias58spin, anderson59theory}. Thus, such a conventional superconducting state with a spin-singlet pairing is suppressed or destroyed by time-reversal symmetry breaking perturbations — as a consequence of applied magnetic field, doped magnetic impurities, or intrinsic magnetic instability leading to spontaneous magnetization — due to electron pair breaking. 

On the other hand, beyond conventional BCS paradigm, unconventional superconductivity allows coexistence of more exotic superconducting order parameters with magnetic order. For instance, as predicted independently by Peter Fulde and Richard Ferrell (FF) \cite{FF64} and Anatoly Larkin and Yuri Ovchinnikov (LO) \cite{LO65}, magnetic fields can give rise to a superconducting state with FF-type order parameter $\Delta(x)= \Delta e^{iqx}$ and/or spatially inhomogeneous LO-type pair potential $\Delta(x)= \Delta \cos qx$. The underlying physical mechanism of the Fulde–Ferrell–Larkin–Ovchinnikov (FFLO) state \cite{FF64,LO65}, owning to the opposite energy-shift in the electronic spin bands as shown in Fig. \ref{HSC}(B), induces non-zero centre-of-mass momentum of Cooper pairs and leads to a spatially-modulated order parameter. The FF state ubiquitously exist in noncentrosymmetric SCs and is particularly known as the helical superconductivity \cite{bauer2012,TSC-RPP,agterberg03,Barzykin02,dimitrova03, Kaur05,Agterberg07,Dimitrova07,Samokhin08,yanase08,Michaeli12, Sekihara13,Houzet15}.

The FFLO states, and/or the implications of the helical superconductivity, have been obtained in heavy-fermion SCs CeCoIn$_5$ \cite{Bianchi03,radovan03,Matsuda07Rev}, organic SCs \cite{Wosnitza18Rev}, pure single crystals of FeSe \cite{Shigeru14,Kasahara20}, thin films of Pb \cite{Sekihara13} and doped SrTiO$_3$ \cite{Schumann20}, a heavy-fermion Kondo superlattice \cite{Naritsuka17,naritsuka21Rev}, and a three-dimensional topological insulator Bi$_2$Se$_3$ \cite{chen18-TI}. While the existence of FFLO-like states is well established in proximity-coupled SCs and ferromagnets \cite{Buzdin05RMP}, the experimental observation of FFLO states has been reported in nonmagnetic SCs by applying external magnetic fields \cite{Bianchi03,radovan03} as well as intrinsic ferromagnetic SCs \cite{Felner97,aoki01,pfleiderer01,Ren09, Dikin11,li11,bert11}.

\subsubsection{Pairing in Rashba/Ising SCs}
To understand SDE via finite-momentum Cooper pairing in noncentrosymmetric superconducting materials, it is instructive to quickly review pairing phenomenon in Ising- and Rashba-type superconductivity. In this regard, (5QL)Bi$_2$Se$_3$/NbSe$_2$(ML) heterostructure is a promising example where a crossover from Ising- to Rashba-type superconductivity is reported recently \cite{yi22crossover}. NbSe$_2$ bulk crystal with 2H phase is a well-studied superconductor with Fermi surface sheet-dependent s-wave superconductivity \cite{Yokoya01}. 2H-NbSe$_2$ bulk crystals covered by molecular-beam epitaxy (MBE)-grown films of Bi$_2$Se$_3$ or Bi$_2$Te$_3$ topological insulators are the most successful topological superconductor interfaces \cite{TSC-Wang,xu14momentum, xu14artificial, xu15experimental,sun16majorana, zhu21discovery}. It is also well-known that monolayer NbSe$_2$ with the Se-Nb-Se trilayer structure, with preserved out-of-plane mirror symmetry but broken in-plane inversion symmetry, is a prototypical Ising-type superconductor \cite{Xi16,Xing17}, and are preferred over 2H-NbSe$_2$ bulk crystals for for device fabrication and technological applications. On the other hand, Bi$_2$Se$_3$ with the Se-Bi-Se-Bi-Se Quintuple-layered structure is a prototypical 3D strong TI hosting a single surface Dirac cone intertwined with nontrivial bulk Rashba bands at the $\Gamma$-point \cite{zhang09,xia09}.\par

In monolayer NbSe$_2$, broken in-plane inversion symmetry generates an out-of-plane spin polarization and originates the Ising-type SOI which induces a valley-dependent Zeeman-type spin-splitting, as shown in figure \ref{HSC}(A-a). Such opposite spin-splitting in the bulk valence bands around valleys K and K$^\prime$ leads to Ising-type superconducting pairing symmetry in monolayer NbSe$_2$, i.e., which refer to the intervalley spin-momentum locked spin-singlet Cooper pairing between two electrons with opposite momenta and opposite out-of-plane spins. On the other hand, as shown in figure \ref{HSC}(A-b), owning to the emergence of Rashba-split low-energy conduction bands at $\Gamma$-point and corresponding Dirac surface states, (5QL)Bi$_2$Se$_3$/NbSe$_2$(ML) heterostructures become proximity-coupled topological SCs with Rashba-type superconducting pairing symmetry, i.e., which refer to the Cooper pairing between two Rashba-split electrons with opposite momenta and opposite spins pinned to the in-plane direction.\par

\subsubsection{Nonreciprocity in FFLO states}
Such momentum-dependent spin-splitting of the low-energy electronic bands, caused by broken inversion symmetry in the noncentrosymmetric bulk crystals, surfaces, and interfaces, is crucial for the emergence of nonreciprocal transport. However, in order to avoid the cancellation of this effect due to superposition of degenerate Kramers pairs, one also needs energy-dependent spin-splitting such that electrons with opposite momenta and opposite spin become Kramers non-degenerate, or simply non-equivalent. Typically, this can achieve by breaking time-reversal symmetry. In the presence of external magnetic field or intrinsic magnetization, parallel to the electron's spin-orientation, BSC-type zero-momentum Cooper pairing (symmetric around $\Gamma$-point) become asymmetric around $\Gamma$-point due to opposite energy-shift and FFLO-type finite-momentum Cooper pairing originates in both Ising- and Rashba-type superconducting phase.\par

Recent theoretical studies \cite{Daido22,Noah22,He22,Ilic22} revealed how a non-trivial interplay of antisymmetric Rashba SOI, magnetic field, and helical supercurrent leads to an intrinsic SDE in noncentrosymmetric bulk SCs. It implies that intrinsic SDE is closely related to the FFLO state \cite{FF64,LO65} with a periodically modulating phase of the superconducting order parameter $\Delta_{sc}(r)=\Delta_{sc}e^{iq\cdot r}$: Rashba SOI splits Fermi surfaces while the finite pairing momentum $\textbf{q}_0$ is induced by the magnetic field and varies continuously with its strength and orientation. In terms of charge transport, when an in-plane magnetic field is applied, Cooper pairs in noncentrosymmetric Rashba SCs acquire a finite-momentum $\textbf{q}_0$, and, as a result, critical currents traversing along the direction parallel and antiparallel to $\textbf{q}_0$ become unequal. \par

Recently, N. Yuan and L. Fu \cite{Noah22} explicitly demonstrated the effect of in-plane magnetic field on Rashba spin-split bands and, thus, the emergence of finite-momentum Cooper pairing. As shown in figure \ref{HSC} (D-a) and (D-c), a finite magnetic field (B) displaces the centers of Rashba-split inner(+) and outer(-) Fermi pockets from $\textbf{k}=0$ to opposite momenta, $\pm\textbf{k}_0=\pm \hat{z}\times B/v_F$, respectively, and leads to a finite intrapocket Cooper pair momentum $\textbf{q}_0=\pm2\textbf{k}_0$. Owning to the larger DOS in the outer pocket, usually, energetically favored state is the one with the Cooper pair momentum $\textbf{q}_0=-2\textbf{k}_0$. Figure \ref{HSC} (BR-b) and (BR-d) show a magnetic field dependence of the depairing critical current in the fluctuation regime of a metal-superconductor resistive transition. When $\textbf{B}\parallel \textbf{J}$, as shown in figure \ref{HSC} (BR-b), the phase diagram in the B-J plane remains symmetric with respect to both B and J axes and thus, no nonreciprocity in the critical current $J_c$ and critical magnetic field $B_c$. However, when $\textbf{B}\perp \textbf{J}$, as shown in figure \ref{HSC} (BR-d), the phase diagram becomes asymmetric/skewed, indicating nonreciprocity in the critical current $J^+_c\ne J^-_c$ and polarity-dependence of critical field $B^+_c\ne B^-_c$. That is, the maximum critical current flowing in the direction parallel and antiparallel to $\textbf{q}_0$ are different, which leads to SDE. On the same footing, in the presence of a supercurrent, the polarity-dependence of in-plane critical fields is also a direct consequence of the finite-momentum Cooper pairing. Similar mechanism has been realized in Rashba SCs with intrinsic magnetization. Figure \ref{HSC}(BL) demonstrates Rashba spin-splitting and SDE by controlling magnetization orientation in a JJ Nb/Pt/Nb with a proximity-magnetized Pt barrier (Pt/Y$_3$Fe$_5$O$_{12}$ (YIG)) \cite{Jeon22}.\par

\subsubsection{From spin-singlet to spin-triplet pairing}
It is also crucial to consider competition between spin-singlet and spin-triplet pairing. Note that, in the absence of magnetic field, pairing momentum remains zero for spin-singlet symmetry even in the presence of SOI, whereas SOI induces finite momentum for spin-triplet symmetry, $q^{\pm}=q^{\pm}_0$. However, owning to the symmetric shift of pairing momentum ($q^+_0=q^-_0$) in the absence of magnetic field, even finite-momentum of spin-triplet pairing does not induce nonreciprocity of supercurrent. It can be explained by noticing that the q-linear term in the kinetic energy of Cooper pairs 
could only shift the momentum space positions of the optimal critical currents (maximum $I^+$ and minimum $I^-$) while keeping $I^{\pm}$ values unchanged, and thus, could not induced nonreciprocity. To induce nonreciprocity, one needs magnetic field dependent (higher order) q-terms in the GL free energy of a SC \cite{He22}.  

When magnetic field is applied, energy-dependent spin-splitting lifts the Kramers degeneracy, such that electrons with opposite spin are not momentum-symmetric around $\Gamma=0$, and nonrecirpocity emerges in both cases. As a result, contribution from Cooper pairs with opposite center of mass momentum becomes non-equivalent and the cancellation of their effect is avoided. In the spin-singlet pairing symmetry, magnetic field changes both the magnitude and the momentum-space position of center of mass momentum: enlarging and moving $q^+_0$ along the current direction while reducing and moving $q^-_0$ opposite to the current direction. On the other hand, with the spin-triplet pairing symmetry, SOI shifts momentum of Cooper pairs from $q_0=0$ (symmetric around $\Gamma$-point) to $q=q^{\pm}_0$ (asymmetric around $\Gamma$-point) due to opposite momentum-shift. Unlike spin-singlet case, magnetic field cannot change the momentum space position of $q=q^{\pm}_0$ but rather modifies their magnitude: $q^+_0$ enlarges whereas $q^-_0$ reduces with increasing magnetic field. 

There is a threshold limit, certainly, for magnetic field. For the spin-triplet pairing, when bottom of one CB passes above FL, $q^-_0\rightarrow0$ while $q^+_0$ become maximal. On the other hand, for the spin-singlet pairing, magnitude of $q^-_0$ become constant when in the inner Fermi circle shifted completely on one-side of $\Gamma$-point. Furthermore, one needs to keep an eye on the curvature of parabolic bands as it is key to understand change in momentum when magnetic field is increased, means the Fermi velocity plays central role.

\section{\label{Theory}Theory of superconducting diode effects}
Before jumping onto the recently reported theoretical analysis of SDE, it is important to have a quick review of theoretical studies in which nonreciprocity of supercurrent is reported and the interesting functionalities of SCs intertwined with broken inversion symmetry and SOI are highlighted. Interestingly, V. M. Edelstein has discussed the characteristics of the Cooper pairing in two-dimensional noncentrosymmetric electron systems \cite{Edel89CP}, magnetoelectric effect in polar SCs \cite{Edel95MCA}, and nonreciprocity in the supercurrent by studying the Ginzburg-Landau equation for SCs of polar symmetry \cite{Edel96GL}. In other words, SDE has been there since 1990s, and only recently demonstrated experimentally. \par

In 1996, followed by his earlier work characterizing Cooper pairing in noncentrosymmetric SCs \cite{Edel89CP} and describing magnetoelectric effects polar SCs \cite{Edel95MCA}, V. M. Edelstein \cite{Edel96GL} explicitly proposed nonreciprocity in the supercurrent, that is, when applied magnetic field ($B$), electric current ($j$), and polar axis ($\hat r$) are orthogonal to each other, the magnitude of the critical current $j_c(B)$ depends on the sign of the mixed product $(\hat r\times \hat B)\cdot \hat j_c$, i.e., the critical current should be different for two opposite directions.

Recently, intriguing experimental demonstrations of SDE, especially in the Rashba-type bulk superconducting [V/Nb/Ta]$_n$ superlattice \cite{Ando20} or Al/InAs-2DEG/Al JJs \cite{Baumgartner22} and in the Ising-type superconducting JJs such as NbSe$_2$ constriction \cite{Lorenz22} or Nb/NiTe$_2$/Nb junction \cite{Pal22}, has stimulated theoretical research on nonreciprocal supercurrent transport in a number of exotic quantum materials. In addition, it also sparked the discussion on fundamental mechanisms that cause nonreciprocal charge transport in SCs. For instance, how nonracirocal charge transport in a semiconductor with finite resistance could be generalized to a superconductor allowing supercurrent with zero-resistance? More specifically, which physical quantity display nonreciprocal behaviour in the \par

By employing mean-field (MF), Bogoliubov–de Gennes (BdG), and time-dependent Ginzburg-Landau (GL) theories, Daido et al. \cite{Daido22}, J. He et al. \cite{He22}, N. Yuan and L. Fu \cite{Noah22}, and S. Ilić and F. S. Bergeret \cite{Ilic22} theorized SDE in junction-free Rashba/polar SCs. A. Daido et al. \cite{Daido22} studied Rashba-Zeeman-Hubbard model for the helical superconductivity and proposed that nonreciprocity in the depairing critical current is the intrinsic mechanism of SDE in the fluctuation regime of metal-superconductor resistive transition. A. Daido et al \cite{Daido22} also showed that such mechanism of intrinsic SDE can be employed as a microscopic probe to study and explore the phase diagram of helical superconductivity. Similar proposal has been made by N. Yuan and L. Fu \cite{Noah22}, who studied effective Rashba-Zeeman-Hubbard model and reported that nonreciprocal depairing critical current and the polarity-dependent critical magnetic field are the consequences of finite-momentum Cooper pairing. On the same footing, mainly using the GL theory and phenomenological theory of SDE, J. He et al. \cite{He22} presented a detailed discussion on symmetry breaking phenomenon and an intertwining between polar axis, magnetic field orientation, and current direction that is desired for the realization of SDE. The theory of SDE has been generalized for Rashba SCs with arbitrary disorder by S. Ilić and F. S. Bergeret \cite{Ilic22}.\par

Thus far, theoretical discussion on nonreciprocal supercurrent and prediction of intrinsic SDE has also been extended for other junction-free polar superconducting systems. For instance, H. D. Scammell et al. presented theory of zero-field SDE in twisted trilayer graphene \cite{Scammell22}. Zhai et al. \cite{Zhai22} predicted reversible SDE in ferroelectric SCs. The experimental demonstration of nonreciprocal transport in chiral SCs, e.g., Ru-Sr$_2$RuO$_4$ eutectic system \cite{Hooper04,Kaneyasu10} and WS$_2$ nanotubes \cite{Qin17}, is recently followed by B. Zinkl et al. \cite{Zinkl22} who discussed the detailed symmetry conditions for the SDE in various chiral superconducting models/systems. The theory of nonreciprocal charge transport and intertwining between SDE and band topology has also been presented for topological SCs \cite{Noah21,Legg22,Takasan22}. For instance, N. Yuan and L. Fu \cite{Noah21} uncovered an intertwining between finite-momentum superconductivity and topological band theory, i.e., Cooper pairing with finite momentum depends closely on the nontrivial topological spin texture of nondegenerate Fermi surfaces, driven by combined effect of SOI and Zeeman fields. Recently, H. F. Legg et al. \cite{Legg22} theorized SDE due to MCA in topological insulators and Rashba nanowires, while K. Takasan et al. \cite{Takasan22} discussed supercurrent-induced topological phase transitions.\par

In addition, the basic mechanisms of SDE (first envisioned by J. Hu et al. \cite{Hu07}) has also been theorized for JJs \cite{Davydova22}, e.g., conventional superconducting NbSe$_2$/Nb$_3$Br$_8$/NbSe$_2$ JJ \cite{Zhang22-JJ} and Al/InAs-2DEG/Al JJ \cite{Baumgartner22-JPCM}, graphene-based JJ \cite{Wei22}, and topological superconducting JJ \cite{Pekerten22,Tanaka22d-wave}. Furthermore, the effect of Rashba and Dresselhaus SOI on supercurrent rectification and MCA have also been studied for JJs based on conventional SCs \cite{Baumgartner22-JPCM} topological SCs \cite{Pekerten22}. In the recent theoretical studies on the topological JJ dS/FI/dS (dS: d-wave superconductor, FI: ferromagnetic insulator) on a 3D topological insulator surface, Y. Tanaka and N. Nagaosa \cite{Tanaka22d-wave} also demonstrated the relevance of the Majorana bound states (MBS), i.e., spin-momentum locked energy-zero Andreev bound states (ABS) at the interface \cite{FuKane08,Linder10}. \par

\section{\label{Materials}Materials for superconducting diode effects}
In the last two years, SDE has been experimentally observed in a number of superconducting structures, ranging from junction-free SCs \cite{Ando20,Miyasaka21,Narita22,Kenji19,masuko22,Lin22}, JJs \cite{Baumgartner22,Jeon22,Golod22,Wu22,Lorenz22,Pal22,DezMrida21}, and other engineered structures such as superconducting tunnelling junctions \cite{Strambini22} and superconducting devices with pinning centres of asymmetric pattern \cite{Lyu22}. JJs, mainly due to the presence of a junction, can be though as symmetric and superconducting analogue of asymmetric semiconducting pn junction. On the other hand, junction-free SCs can be though as symmetric and superconducting analogue of symmetric semiconductors.\par

The observation of SDE originated from nonreciprocal charge transport driven by MCA in symmetric SCs , whether junction-free or JJs, relies on simultaneously broken spatial-inversion and time-reversal symmetries, similar to that in symmetric semiconductors. Furthermore, similar that in topologically nontrivial semiconductor/semimetals, SDE can be realized in time-reversal symmetric systems where nonreciprocal charge transport is associated with nontrivial Berry phase. Since both MCA and nontrivial Berry phase are strongly associated the strength and nature of SOI originated due to broken inversion symmetry, noncentrosymmetric SCs can be classified as Rashba SCs \cite{Ando20,Miyasaka21,Narita22,Kenji19,masuko22,Baumgartner22,Jeon22,Golod22} or Ising SCs \cite{Wu22,Lorenz22,Pal22}. 

If spatial-inversion symmetry is broken, SDE can be realized in three-dimensional bulk materials, quasi-two-dimensional thin films and van der Waals heterostructures, and atomically-thin superconducting materials. Thus far, SDE has been reported in several materials, ranging from conventional SCs such as [Nb/V/Ta]$_n$ superlattice \cite{Ando20,Miyasaka21}, Al/InAs-2DEG/Al junction \cite{Baumgartner22}, Nb SCs \cite{Golod22}, Cu/EuS/Al tunnel junction \cite{Strambini22}, and superconducting thin films with conformal-mapped nanoholes \cite{Lyu22}, ferromagnetic SCs \cite{Narita22}, twisted-angle bilayer \cite{DezMrida21} and trilayer \cite{Lin22} graphene with unconventional superconductivity and, TMDCs with Ising superconductivity \cite{Wu22,Lorenz22,Pal22}, and topological superconducting materials \cite{Kenji19,masuko22,Pal22} where superconductivity coexists with nontrivial band topology.

For device fabrication of superconducting electronics, and especially for the search/utilization of novel superconducting materials with high workable temperature and large magnetic field, it is important to categorize materials hosting SDE. Superconducting materials/structures displaying SDE can be classified as junction-free or JJs based on device structure, Rashba or Ising SCs based on the nature of SOI, and trivial or nontrivial based on band topology. Furthermore, SDE can be classified as magnetic-field-driven or field-free SDE depending on the magnetic character of superconducting materials. Furthermore, depending upon the origin of nanoraciprocity of charge transport, whether MCA or nontrivial Berry phase, SDE materials can be classified as time-reversal-symmetric or time-reversal-asymmetric.\par   

\section{\label{Efficiency}Efficiency of superconducting diode}
Let's consider a superconducting sheet with pairing potential $\Delta(q)$, where $\textbf{q}=q\hat x$ is the center-of-mass momentum. The metal-superconducting transition, and thus a distinction between supercurrent, depairing current, and a normal current, can be conveniently described by introducing condensation energy $F(q)\equiv F_n(q)-F_s(q)$ for each $q$, i.e., the difference between free energy per unit area in the normal (n) and superconducting (s) states. The sheet current density, as an expectation value of the current operator, can be obtained by $j(q)=2\partial_qF(q)$. If a current source supplies an electric current $j_{ex}$, a superconducting state with pairing momentum \textbf{q} should be realized when $j_{ex}=j(q)$. However, when $j_{ex}<j^-_{c}\equiv$ min$_q j(q)$ or $j_{ex}>j^+_{c}\equiv$ max$_q j(q)$, the superconducting state can not sustain $j_{ex}$ and turns into a normal state. Thus, the depairing critical current along a direction parallel ($+\hat x$) and antiparallel ($-\hat x$) to the pairing momentum \textbf{q} is given by the maximum ($j^+_{c}$) and minimum ($j^-_{c}$) value of $j(q)$. The SDE in such helical superconductor is identified and characterized with a finite $\Delta j_c$ given by
\begin{equation}
\Delta j_c\equiv j^+_{c}+j^-_{c}=j^+_{c}-|j^-_{c}| \;\label{DEP}
\end{equation}
Although a huge current density is generally required to achieve the depairing limit in a typical superconductor, depairing critical current density ($j_c$) has recently been reported in the superconducting microbridge devices \cite{Nawaz13,Li13,Sun20}. For an optimal performance of SDE, it is instructive to analyse the behavior of depairing $j_c$ and $\Delta j_c(T)$ through various perspectives. For instance, dependence of critical current density $j_c$ on temperature and the orientation of magnetic field reported for Fe-based Ba$_{0.5}$K$_{0.5}$Fe$_2$As$_2$ microbridge with nanoscale thickness, see, e.g. Fig. 3 and Fig. 4 in ref. \cite{Li13}, critical current density as a function of bridge width and length reported for Cu-based YBa$_2$Cu$_3$O$_{7-\delta}$ microbridge with nanoscale thickness, see, e.g. Fig. 3 in ref. \cite{Nawaz13}, a comparison of critical current density obtained from Ginzburg-Landau (GL) theory $[\propto (T_c-T)^{3/2}]$ to that from Kupriyanov-Lukichev (KL) theory for Fe-based Fe$_{1+y}$Te$_{1-x}$Se$_x$ microbridge with microscale thickness, see figure 3 in ref. \cite{Sun20}, and the sign reversal of $\Delta j_c$ by increasing the magnetic field at low temperatures, see, e.g. Fig. 4 and Fig. 5 in ref. \cite{Daido22}. As a figure of merit, the strength of the nonreciprocal response or the superconducting diode efficiency can be expressed as a ratio between $\Delta j_c$ and the averaged critical current $j^{avg}_c$ \cite{Daido22,He22,Noah22,Ilic22}
\begin{equation}
\eta\equiv\frac{j^+_c-|j^-_c|}{j^+_c+|j^-_c|}=\frac{\Delta j_c}{2j^{avg}_c}  \;\label{Eff}
\end{equation}
Recent theoretical studies \cite{Daido22,He22,Noah22,Ilic22} show that the strength of $\eta$ depends on a range of relevant system parameters: applied magnetic field, working temperature, induced Cooper pairing momentum, intrinsic SOI, and an intertwining between them \cite{Daido22,He22,Noah22,Ilic22}. In addition, strength of $\eta$ also depends on two other related but distinct parameters, chemical potential \cite{He22} and next-nearest neighbour hopping \cite{Daido22} that break the particle-hole symmetry. Furthermore, though the SDE persists even in the presence of disorder, strength of $\eta$ is also affected by disorder as it may cause changes in the nature of the two helical bands by introducing mixing between them \cite{Ilic22}. Thus, for energy-efficient and high performance superconducting device application, it is crucial to find certain optimal system parameter regimes where the strength of $\eta$ is maximal. Recently, Ilice et al. \cite{Ilic22} theoretically predicted that SDE efficiency may exceed $\eta=40\%$ (in the ballistic limit) at optimal magnetic field, temperature, and SOI in Rashba SCs. Interestingly, SDE with optimal efficiency can be engineered by steering the exotic characteristics and the design of a JJ \cite{Baumgartner22,Tanaka22d-wave}.\par

\begin{figure*}
\includegraphics[scale=0.4]{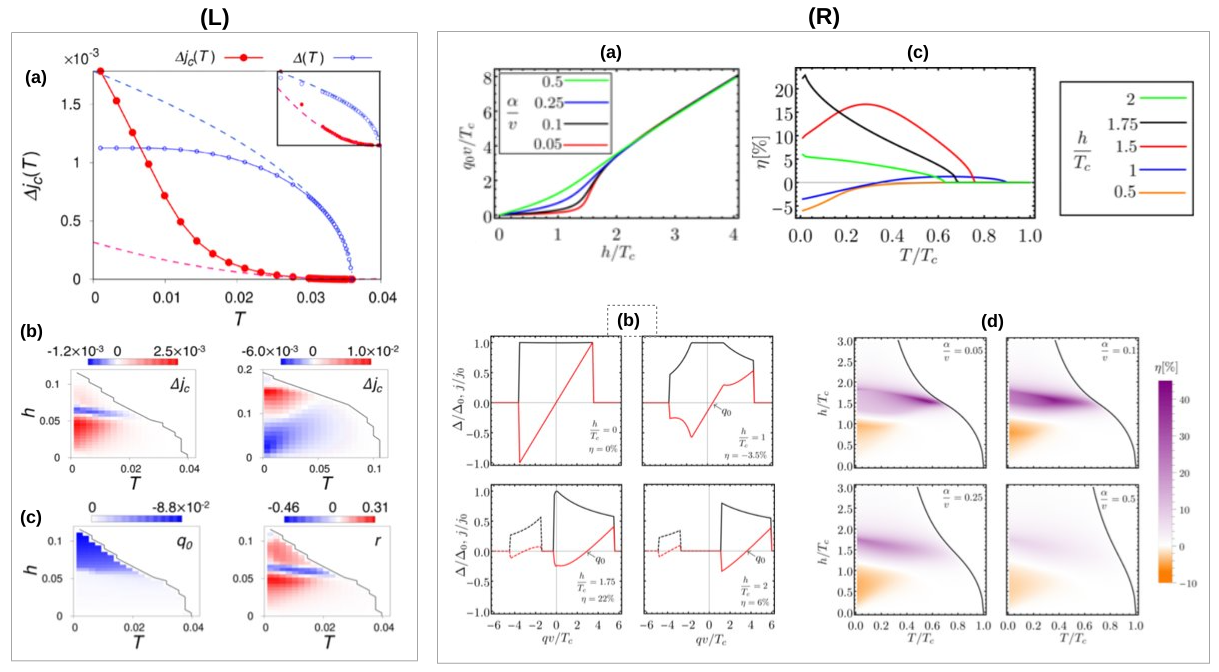}
\caption{\label{DE}\textbf{Dependence of superconducting diode effect on the system parameters and its optimization} \textbf{(L)} Rashba-Zeeman-Hubbard model for a Rashba superconductor. \textbf{(a)} The temperature dependence of $\Delta j_c(T)$ (red closed circles) and $\Delta(T)$ (open blue circles) with arb. units. The red and blue dashed lines represent the fitting curves of $\Delta j_c(T)$ (with $(T_c-T)^2$) and $\Delta(T)$ (with $\sqrt{T_c-T}$) respectively. Inset: Enlarged view near $T_c$. Here Zeeman exchange parameter is set as $h = 0.03$ and the transition temperature reads $T_c\approx0.036$ respectively. \textbf{(b)} The h-T phase diagram depicting temperature and the magnetic field dependence of $\Delta j_c(h,T)$ with $t_2=0$ (left) and with $t_2=0.2$ (right). Here $t_2$ denotes next-nearest-neighbour hopping while the red (blue) color indicates positive (negative) values of $\Delta j_c$. \textbf{(c)} Pairing momentum $q_0$ (left) and SDE efficiency (right), represented by $r$ here, for various values of h and T. Figure (L) is reproduced with permission from ref. \cite{Daido22}. \textbf{(R)} Quasiclassical Eilenberger equation for a 2D Rashba superconductor \textbf{(a)} Helical modulation vector $q_0$ as a function of magnetic field, where $q_0v\approx2(\alpha/v)h$ corresponding to the “weak helical phase” at low fields, whereas $q_0v\approx2h$ corresponding to the “strong helical phase” at high fields. Here $q_0$ is calculated in the vicinity of the upper critical field ($h_{c2}$) at different strengths of spin-orbit interaction ($\alpha/v$). \textbf{(b)} Supercurrent $j$ (red) and the superconducting gap $\Delta$ (black) plotted as a function of the phase gradient, for different magnetic fields at fixed values of temperature $T=0.01T_c$ and spin-orbit interaction $\alpha/v=0.25$. Both quantities are calculated self-consistently and the curves are normalized with $j_0$ and $\Delta_0$, respectively, which represent the critical current and the superconducting gap at $T=h=0$. \textbf{(c)} Temperature-dependence of superconducting diode efficiency $\eta$, calculated for different values of the magnetic field and fixed spin-orbit interaction $\alpha/v=0.25$. \textbf{(d)} Superconducting diode efficiency $\eta$, calculated for different strengths of spin-orbit interaction in the ballistic limit, corresponding to every point in the h-T phase diagram. Here black curve corresponds to the upper critical field $h_{c2}$ while the purple (orange) color indicates positive (negative) values of $\eta$. Figure is reproduced with permission from ref. \cite{Ilic22}.}
\end{figure*}

At some fixed temperature, SDE efficiency shows nonmonotonic magnetic field dependence \cite{Daido22,Noah22,Ilic22}: $\eta$ increases (almost linearly) for (weak) moderate fields and then suppresses beyond a certain breakdown/threshold field $B_{max,\eta}$, see, e.g. Fig. 3(D) in ref. \cite{Noah22} and Fig. 4 in ref. \cite{Ilic22}. For Rashba SCs, threshold field is theoretically \cite{Noah22,Ilic22} predicted to be of the order of the Pauli paramagnetic limit, i.e., much larger than the breakdown limit observed in recent experiments \cite{Ando20,Baumgartner22,Lorenz22}. Along with this nonmonotonic behavior, SDE efficiency changes its sign with increase in magnetic field \cite{Daido22,Noah22,Ilic22}. Such change in sign of SDE efficiency appears approximately at the Pauli limit $B\approx B_P$, see, e.g. Fig. 3(F) in ref. \cite{Noah22}, Fig. 3 in ref. \cite{Ilic22}, and Fig. 4 in ref \cite{Daido22}. Such magnetic field driven sign reversal of the SDE, accompanied by the crossover between weak and strong helical phase, is a general feature of helical SCs irrespective of their details \cite{Daido22-PRB}.\par

Unlike magnetic field dependence, recent theoretical studies predict quite diverse behaviour for the temperature dependence of SDE efficiency. For instance, at some fixed magnetic field, Rashba-Zeeman-Hubbard model \cite{Daido22,He22,Noah22} predict that SDE efficiency shows a monotonic square-root-like temperature dependence near the transition temperature which saturates at low temperatures, see, e.g. Fig. 2 in ref. \cite{Daido22}, and Fig. 3 in ref \cite{He22}. On the other hand, quasiclassical Eilenberger equation for a 2D disordered Rashba superconductor \cite{Houzet15} shows that the temperature dependence of SDE efficiency is critically affected by the strength of fixed magnetic field and may display nonmonotonic temperature-dependence \cite{Ilic22}. For instance, SDE efficiency shows a monotonic temperature-dependence for $B\gtrapprox B_P$ but it becomes nonmonotonic when $B\lessapprox B_P$, see, e.g. Fig. 4 in ref. \cite{Ilic22}. That is, in the later case, first SDE efficiency increases with decrease in temperature but it is gradually suppressed when temperature is further lowers after certain breakdown limit. Recent observation of SDE shows that the monotonic \cite{Baumgartner22,Lorenz22} and the nonmonotonic \cite{Lorenz22} temperature-dependence of SDE efficiency may also depend on the sample fabrication \cite{Lorenz22}. Similar transition from monotonic to nonmonotonic temperature dependence may also be realized by varying strength of disorder, see, e.g. Fig. 6 in ref. \cite{Ilic22}\par.  

Next, we turn to the dependence of SDE efficiency on the momentum of Cooper pairs or the nature of helical phase. For spin-orbit coupled Rashba SCs in magnetic field, the nature of helical phase can be characterized by quantifying the contribution of two helical bands to the helical superconductivity \cite{Houzet15}. Owning to the opposite energy shift induced by magnetic field, the two helical bands denoted with the index $\lambda=\pm$ and characterized by the same Fermi velocity $v=\sqrt{2\mu/m+\alpha^2}$ but different densities of states $\nu_\lambda=\nu(1-\lambda\alpha/v)$, prefer opposite modulation vectors: $q^{\lambda}_0v=-2\lambda h$. Here, $m$ is the effective electron mass, $\mu$ is the chemical potential, $\nu=m/(2\pi)$, and $\alpha=\Delta_{so}/\sqrt{2m\mu}$ characterizes the SOI strength. Figure \ref{DE}(R-a) illustrates the crossover from a “weak” to “strong” helical phase for different ratios of Fermi velocity ($v$) and the velocity associated with Rashba SOI ($\alpha$). In the “weak” or long-wavelength helical phase subject to low magnetic fields, contribution of both bands to helical superconductivity yields a modulation vector $q_0v\approx2(\alpha/v)h$. In the “strong” or short-wavelength helical phase at large magnetic fields, owning to the dominance (suppression) of contribution from the band with higher (lower) density of states, only one of the bands contributes to helical superconductivity which leads to the modulation vector $q_0v\approx2h$.\par

S. Ilić and F. S. Bergeret \cite{Ilic22}, based on the quasiclassical Eilenberger equation for a 2D Rashba superconductor \cite{Houzet15}, predicted that the maximum of $\eta$ emerges when both the bands contribute to the helical superconductivity and the magnetic field is close to the critical value $h^*$ at which a crossover between “weak” and “strong” helical superconducting phase occurs. It can be explained from the self-consistent calculation of $\Delta(q)$, $j(q)$ and $\eta$ vs Cooper pairing momentum under various magnetic field strength as shown in Fig. \ref{DE}(R-b) or from the h-T phase diagram under various strengths of Rashba SOI as shown in Fig. \ref{DE}(R-d). In the absence of magnetic field ($h=0$), as shown in the upper left panel of Fig. \ref{DE}(R-b), there is no helical phase ($q_0=0$) and thus no nonreciprocity of the critical current. In the presence of finite magnetic field ($h\ne0$), finite Cooper pairing momentum ($q_0\ne0$) leads to nonreciprocity of the critical current in both the “weak” helical state induced by sufficiently low h, as shown in the upper right panel of Fig. \ref{DE}(R-b), and the “strong” helical state induced by large h, as as shown in the two lower panels of Fig. \ref{DE}(R-b). The momentum dependence of $\Delta(q)$ and $j(q)$ is markedly different in these three superconducting states, and thus, depict a completely different supercurrent transport: no SDE in the BCS state ($j^+_c=|j^-_c|, \eta=0$), whereas negative SDE in the “weak” helical state ($j^+_c<|j^-_c|, \eta<0$) while positive SDE in the “strong” helical states ($j^+_c>|j^-_c|, \eta>0$). In addition, different strength and opposite sign of SDE under different magnetic field values hint that there must be some optimal field at which $\eta$ should be maximum.\par

It can be depicted by plotting $\eta$ for every point in the h-T phase diagram, and in addition, effect of other parameters can be visualised. For instance, as shown in Fig. \ref{DE}(R-d), S. Ilić and F. S. Bergeret \cite{Ilic22} plotted the h-T phase diagram and calculated $\eta$ for different strengths of SOI. Here black curve corresponds to the upper critical field $h_{c2}$ while the orange and purple colors clearly illustrate the two distinct regimes in which SDE is driven by the “weak” and “strong” helical phases, respectively. First, it showcases that the maximum efficiency appears at the crossover between “weak” and “strong” helical phases. Second, maximum efficiency exceeding 40$\%$ at the crossover corresponds to the optimal SOI. Third, the  maximum efficiency also corresponds to optimal temperature in the superconducting phase.\par

Such momentum dependence, yielding maximum $\eta$ with optimal magnetic field and SOI driving system at the crossover between “weak” and “strong” helical phases, implies that the competition and the contribution of both helical bands is central for the SDE. This can be explained by noticing that the MCA is proportional to magnetic field and SOI, and thus become strongest when both of these parameters are maximal. The maximal magnetic field and SOI borne by the system, along with the constraint of contribution from both helical bands, is ensured at the crossover between “weak” and “strong” helical phases. This can further be explained by the analysing the h-T phase diagram regimes, as illustrated in Fig. \ref{DE}(R-d), where too large magnetic field and too large SOI both suppress the SDE efficiency. For instance, SDE efficiency vanishes when magnetic field is increased, beyond the crossover to the “strong” phase, where only one of the helical bands dominates. Similarly, when SOI is increased — such that $\alpha/v\rightarrow1$, only one helical band with a large DOS ($\nu_-\approx2\nu$) exists while the helical band with vanishingly small DOS ($\nu_+\rightarrow0$) other is fully suppressed, and the SDE disappears.\par

This phase diagram also helps to understand the intertwining of optimal temperature with magnetic field and SOI. At weak SOI, such as depicted in the upper left panel of Fig. \ref{DE}(R-d), SDE becomes strongest at the tricritical point $(T^*,h^*)$ where the “weak” helical phase meets the “strong” helical phase and the normal phase. It is in good qualitative agreement with the results predicted by N. Yuan and L. Fu \cite{Noah22} where $(T^*,h^*)$ denotes tricritical point at which the FF phase meets the normal phase and the BCS phase. However, with increasing strength of SOI, h-T phase diagram regime hosting maximum SDE moves towards zero-temperature, i.e., where $T\ll T^*$.

Similarly, as shown in Fig. \ref{DE}(L-b), Daido et al. \cite{Daido22} plotted the h-T phase diagram and calculated $\eta$ for different strengths of next-nearest neighbour hopping $t_2$ in the Rashba-Zeeman-Hubbard model. It depicts the sign change of $\eta$ with increasing magnetic field. In addition, at some magnetic field, the sign of SDE efficiency found at $t_2=0$ (left panel) also switches when a finite $t_2\ne0$ is considered (right panel). Furthermore, magnetic field dependence of pairing momentum as shown in the left panel of Fig. \ref{DE}(L-c) and the SDE efficiency as shown in the right panel of Fig. \ref{DE}(L-c) showcases that the maximum $\eta$ appears at the crossover between “weak” and “strong” helical phase. It implies that the results obtained from the numerical study of Rashba-Zeeman-Hubbard model \cite{Daido22} and that from quasiclassical Eilenberger equation \cite{Ilic22} are in good qualitative agreement. However, as mentioned above, there is considerable differences between these two studies when it comes to the temperature dependence of SDE efficiency, i.e., numerical study of Rashba-Zeeman-Hubbard model shows monotonic behaviour while quasiclassical Eilenberger equation shows temperature dependence could be either monotonic or nonmonotonic depending on the strength of magnetic field. Based on the above analysis, one can conclude that the nonmonotonic behaviour, for both magnetic field and temperature dependence, and the change of sign of $\eta$ with increasing magnetic field is related to the magnetic field-driven evolution of the helical phase. That is, $\eta$ becomes maximum at a particular field $h^*$ and optimal temperature, and then lowers for other values.

Similar to the dependence on next-nearest neighbour hopping \cite{Daido22}, and consistent with the analogy discussed for magnetic field and SOI intertwined with the variation in the DOS of two helical bands \cite{Ilic22}, He et. al \cite{He22} theoretically predicted that SDE efficiency show strong dependence on the chemical potential. For a Rashba superconductor with Zeeman field, where the free energy includes all terms up to the linear order in $h\sqrt{\epsilon}$, GL theory results in the SDE efficiency \cite{He22}:
\begin{equation}
\eta=\frac{2.7\lambda_R}{|\lambda_R|}\frac{h\sqrt{\epsilon}}{T_c}\times
\begin{cases}
(1+\tilde{\mu})^{-1/2} & \text{if } \tilde{\mu}>0\\
\frac{8}{7}+\frac{16}{21}\tilde{\mu}+(1+\tilde{\mu})^{-1/2}  & \text{if } -1<\tilde{\mu}<0 \;\label{Q}
\end{cases}
\end{equation}
Here $\lambda_R$ is the Rashba SOI strength, $\epsilon=1-T/T_c$, and $\tilde{\mu}=\mu/E_R$ where $E_R=\frac{1}{2}m\lambda^2_R$ is the energy difference between band crossing point ($\mu=0$) of Rashba-split bands and the conduction band edge ($\mu=-E_R$) and $m$ denotes the effective electron mass. At some fixed magnetic field, temperature, and SOI, SDE efficiency shows maximum strength at $\mu=0$, whereas it decrease when the Fermi level moves away from the band crossing point, either towards the large $\mu$ limit ($\mu\gg E_R$) or towards the conduction band edge ($\mu=-E_R$). It is important to note that there are several constraints, and thus limitations, on these GL theory calculations. For instance, the expression (\ref{Q}) is derived by assuming $|h|\ll Tc\ll E_R$ and treating the problem in the band basis where only the intra-band pairing $\Delta_t$ is considered while the inter-band pairing $\Delta_s$ is neglected. As a consequence of taking the limit $T_c/E_R\rightarrow0$ and neglecting the inter-band pairing, there exists a discontinuity in $\eta$ at $\mu=0$. In addition, owning to the consideration of intra-band pairing only, such a discontinuity appears also due to the flip of spin-momentum locking helicity. However, the features of SDE efficiency obtained numerically from a self-consistent Bogoliubov–de Gennes mean-field Hamiltonian \cite{He22} are in good qualitative agreement with those displayed by SDE efficiency obtained from the analytic generalized GL theory calculations. In addition, the discontinuity of $\eta$ at $\mu=0$ is smoothed out when $T_c/E_R$ is not so small and it shows square root dependence on $\mu$, $\eta \sim \mu^{1/2}$, when $\mu$ is large.

Finally, it is important to emphasise that SDE efficiency also depends upon the characteristics and the design of a JJ \cite{Baumgartner22,Tanaka22d-wave}. In general, with a macroscopic phase difference $\phi$ between two SCs, the standard CPR of the Josephson supercurrent $I(\phi)$ between two SCs is $I(\phi) \sim \sin \phi$. That is, when either space-inversion symmetry or time-reversal symmetry is preserved, purely sinusoidal terms leads to an antisymmetric CPR, $I(\phi)=-I(-\phi)$, and the Josephson current vanishes for $\phi=0$. On the other hand, when both time-reversal symmetry and space-inversion symmetry are simultaneously broken, an anomalous CPR \cite{bezuglyi02,buzdin08,Reynoso08,Tanaka09-d-wave,Linder10,Jun-Feng10,Jun-Feng11,reynoso12,yokoyama13,Yokoyama14,shen14, konschelle15,Lu15d-wave,rasmussen16, szombati16,assouline19,mayer20gate,strambini20} (displaying finite anomalous Josephson current even at zero phase difference) contains cosine terms as well. However, even the presence of such cosine term does not suffice to obtain SDE because it simply introduces an anomalous phase shift in the purely sinusoidal CPR and thus the Josephson inductance remains reciprocal (symmetric across the zero-current). Thus, in order to realize SDE, it is mandatory that an asymmetry is induced in the CPR by higher order phase (especially sine) terms such that the cosine terms are not absorbed in a mere phase shift \cite{Baumgartner22,Tanaka22d-wave}. \par

By fabricating Al/InAs-2DEG/Al ballistic JJs, Baumgartner et al. \cite{Baumgartner22} observed supercurrent rectification. When an in-plane magnetic field is applied perpendicular to the current, Rashba superconducting system shows an anomalous Josephson supercurrent due to even (cosine) terms in the CPR \cite{rasmussen16}. Such anomalous CPR contains higher harmonic sine terms if the junction transparency is high \cite{mayer20gate,Baumgartner21}, and thus leads to SDE. By theoretical studying a JJ dS/FI/dS made with d-wave SCs (dS) and a ferromagnetic insulator (FI) on the surface of a 3D topological insulator, Y. Tanaka and N. Nagaosa \cite{Tanaka22d-wave} showed that asymmetric CPR containing a wide variety of phase terms leads to high quality SDE \cite{Tanaka22d-wave}. Apart from the conventional $\sin\phi$ phase term in the Josephson current, energy-zero Andreev bound state (ABS) at the dS/FI/dS interface enhances the $\sin2\phi$ component of $I(\phi)$ \cite{yip93d-wave,Tanaka96d-wave}. When the junction dS/FI/dS is placed on the surface of topological insulator \cite{Linder10d-wave}, simultaneous space-inversion and time-reversal symmetry breaking allows a $\cos\phi$ phase term \cite{Tanaka09-d-wave,Linder10} leading to an exotic current-phase relation with $I(\phi) \ne -I(-\phi)$ \cite{Lu15d-wave} while the energy-zero ABS become MBS due to the spin-momentum locking \cite{FuKane08,Linder10}. The simultaneous existence, with almost the same order, of $\sin\phi$, $\cos\phi$, and $\sin2\phi$ phase terms promises a maximum value of SDE efficiency ($\eta=\pm2$) for the d-wave SCs junction on the surface of topological insulator \cite{Tanaka22d-wave}. In light of this, optimal supercurrent rectification effect of a JJ can be realized by exploiting exotic characteristics of unconventional SCs as well as optimizing junction transparency. \par

\begin{figure*}
\includegraphics[scale=0.4]{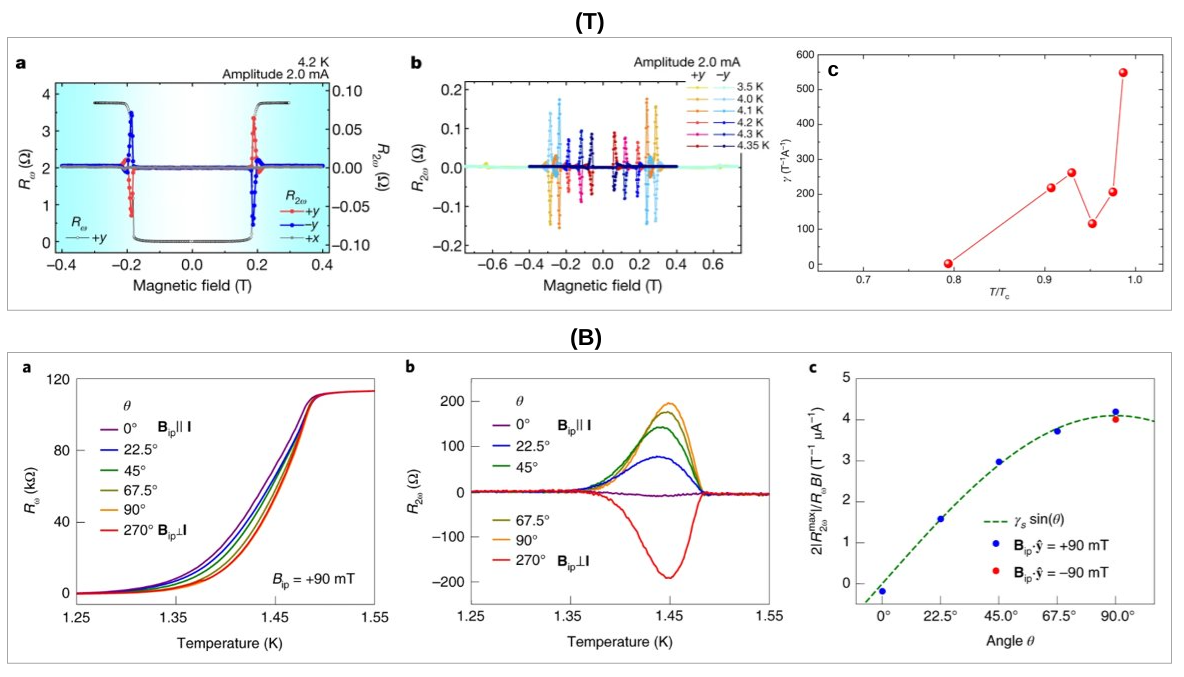}
\caption{\label{RI} \textbf{Magnetochiral anisotropy of the resistance.} \textbf{(T)} Nonreciprocal transport measurements of critical current in the resistive fluctuation regime of [Nb/V/Ta]$_n$ superlattice. \textbf{a} Magnetic field dependence of first-harmonic ($R_{\omega}$) and second-harmonic ($R_{2\omega}$) sheet resistances. $R_{\omega}$ vanishes in the superconducting region (white shadings) while become finite in the normal conducting region (blue shadings). $R_{2\omega}$ enhances when the magnetic field orientation is orthogonal to the current direction and becomes maximal in the fluctuation region. \textbf{b} Temperature dependence of second-harmonic sheet resistance. \textbf{c} Temperature dependence of the coefficient of magnetochiral anisotropy ($\gamma$) calculated from $R_{2\omega}/R_\omega$. The plot roughly shows that $\gamma$ increases with temperature and become maximal in the vicinity of $T_c$, except a a dip appearing at 4.2 K and 4.3 K reflecting small $R_{2\omega}$ values at these temperatures. Figure is reproduced with permission from ref.\cite{Ando20}. \textbf{(B)} Nonreciprocal transport measurements of critical current in the resistive fluctuation regime of Rashba-type Al/InAs-2DEG/Al JJ array. \textbf{(a)} Temperature dependence of first-harmonics $R_{\omega}(T,\theta)$ showing resistive transition for different angles ($\theta$) of the in-plane magnetic field ($B_{ip}$). \textbf{(b)} Temperature dependence of second-harmonics $R_{2\omega}(T,\theta)=V_{2\omega}(T,\theta)/I_{ac}$ of the I-V characteristics for different $\theta$ values with the a.c. current bias of $I_{ac}$= 20 nA. \textbf{c} The coefficient of magnetochiral anisotropy $2R_{2\omega}^{max}/R_{\omega}$ versus orientation/angle $\theta$ of the in-plane magnetic field. Here $R_{2\omega}^{max}$ are the maxima of second-harmonics displayed in (b) and $R_{\omega}$ is the corresponding linear resistance displayed in (a). The red data point shown at $\theta=90^{\circ}$ is obtained by switching the orientation of $B_{ip}$ at $\theta=90^{\circ}$ (the data point in blue), which is equivalent to setting $\theta=270^{\circ}$. The maximal coefficient of magnetochiral anisotropy, extracted from a sine fit of the data, is $\gamma_S\simeq 4.1\times10^6$ T$^-$A$^-$. Figure is reprinted with permission from ref.\cite{Baumgartner22}}
\end{figure*}

\section{\label{Obser}Observation of supercurrent diode effect}
SDE is associated with the literal metal-superconductor transition and defined as nonreciprocity of depairing critical current, i.e., depairing critical current in the direction parallel ($j^+_c$) and antiparallel ($j^-_c$) to the pairing momentum differ ($j^+_c\ne j^-_c$). An ideal SDE would be either $j^+_c$ or $j^-_c$ is zero so that one has maximum $\Delta Jc$. Such a resistive transition between a supercurrent and a normal current can be realized either by extrinsic stimuli or via mechanisms that are intrinsic to the superconducting materials. For instance, the resistive transition can be caused by the vortex motion, usually realized under out-of-plane magnetic fields. Owning to the dependence of dynamics and the statistical mechanics of the vortex system on the device setup such as impurity concentrations and the thermal/quantum fluctuations \cite{RMP94}, such extrinsic mechanism promise tunability of the resistive transition by the nanostructure engineering \cite{Villegas03,Kopasov21}. Apart from the resistance caused by extrinsic mechanisms, the metal-superconductor resistive transition can literally be caused by the dissociation of the Cooper pairs resulting in a transition from supercurrent to a normal current \cite{tinkham04,Dew01}. This occurs at the maximum critical current, which is known as depairing current. In other words, the depairing critical current is directly associated with the closing of the superconducting gap, which reduces and eventually closes with increasing supercurrent. As the depairing limit or the upper limit of the critical current is unique to each superconducting material, depairing current is an intrinsic material parameter for characterizing SCs \cite{RMP94}. Thus, the intrinsic mechanism responsible for SDE ties around the nonreciprocity in the depairing critical current in the fluctuation regime of metal-superconductor resistive transition. In this picture, like many exotic characteristics of quantum materials, intrinsic SDE is a nontrivial quantum mechanical effect. \par

Based on the working temperature, or a working regime of phase diagram representing metal-superconductor resistive transition, observation of SDE can be classified into two main categories: (i) SDE based on the nonreciprocity of depairing current near the superconducting transition temperature ($T\approx T_c$), i.e., in the fluctuation regime of metal-superconductor resistive transition, and (ii) SDE based on the nonreciprocity of supercurrent at sub-Kelvin temperatures ($T\ll T_c$), i.e., deep in the superconducting phase regime. 

\subsection{\label{MCA-R} Magnetochiral anisotropy of the resistance}
In the fluctuation regime of resistive transition close to $T_c$, SDE can be described by MCA of the resistance ($\gamma_S$, as defined in equation (\ref{R})), similar to that in semiconductors, and may be characterized by I-V curves. In this regime, MCA coefficient $\gamma_S$ can be found by measuring the second harmonic signal in lock-in measurements. That is, for an ac current ($I_{in} = I \sin \omega t$) with an amplitude of $I$ and a frequency of $\omega$ applied as input, the nonlinear voltage-drop and current-dependent resistance can be derived from the nonlinear resistance term in equation (\ref{R}) as:
\begin{equation}
\begin{split}
V_{2\omega}(t)&=\gamma BR_\omega I^2\sin^2\omega t\\
&=\frac{1}{2}\gamma BR_\omega I^2\left[1+\sin\left(2\omega t-\frac{\pi}{2}\right) \right]\\
R_{2\omega}&=\frac{1}{2}\gamma BR_\omega I
\end{split}
\end{equation}
Here $R_\omega$ corresponds to the current-independent linear resistance $R_0$, while $R_{2\omega}$ represents the second-order nonlinear resistance, which is dependent on both the current and the magnetic field. Thus by measuring the first- ($R_\omega$) and second-harmonic $R_{2\omega}$ sheet/junction resistances through $2\omega$ voltage response, $\gamma_S$ can be estimated as $\gamma_S=\frac{2R_{2\omega}}{BIR\omega}$.  

However, such resistive measurements cannot realistically simulate the intrinsic SDE at temperatures well below $T_c$ due to no measurable resistance in this regime ($R_0=0$). Thus, the efficiency of SDE is expected to be finite only at $T\approx T_c$ while negligibly small both at temperatures well below Tc and above Tc ($\gamma_N\ll \gamma_S$). For instance, Ando et al. \cite{Ando20} measured MCA of the resistance by performing an a.c. harmonic measurements for Rashba-type bulk superconducting [V/Nb/Ta]$_n$ superlattice \cite{Ando20}. MCA coefficient $\gamma_S$ show sharp increase in the fluctuation regime and reaches to its maximal value $\gamma_S\simeq 550$ T$^-$A$^-$ at $T_c$. However, $\gamma_S$ remains negligibly small at temperatures well below $T_c$. Though the observation seems to be at variance with the theoretical predictions for intrinsic SDE \cite{Daido22,Noah22,He22,Ilic22} and the temperature dependence of experimentally measured MCA in JJs \cite{Baumgartner22,Lorenz22}, but it is an expected outcome of resistive measurements. On the other hand, by fabricating symmetric Rashba-type Al/InAs-2DEG/Al JJs, Baumgartner et al. \cite{Baumgartner22} measured MCA for both the inductance ($\gamma_L$) and the resistance ($\gamma_S$). Finite MCA coefficient $\gamma_S\simeq 4.1\times10^6$ T$^-$A$^-$ observed through resistive measurements near $T_c\sim1.45$ K is of the same order (namely, in the range of $10^6$ T$^-$A$^-$) of the corresponding MCA coefficient observed for the inductance (measured at $T=100$ mK), $\gamma_L\simeq 0.77\times10^6$ T$^-$A$^-$. \par

\begin{figure*}
\includegraphics[scale=0.5]{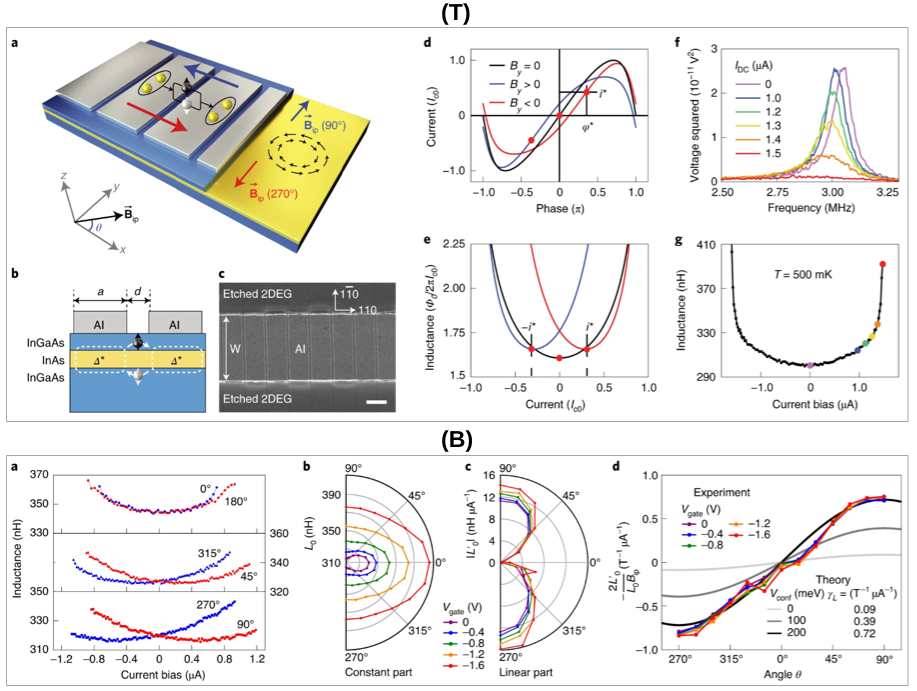}
\caption{\label{LI} \textbf{Current-phase relation and nonreciprocity of inductance in a JJ array.} \textbf{(T)} Device fabrication, current phase relation, and measurement of inductance. \textbf{(a)} A JJ array is made of 2,250 Al islands (grey), of width w=3.15 $\mu$m, length a=1.0 $\mu$m and separated by d=0.1 $\mu$m, on top of a Rashba-type InAs quantum well (yellow) sandwiched between InGaAs barriers. Red and blue arrows represent the spontaneous supercurrents, with zero phase difference, via spin-split pairs of Andreev bound states, denoted by black and white particle representing oppositely spin-polarized electron and hole. The strength and direction of these spontaneous supercurrents depend on that of an in-plane magnetic field $B_{ip}$. Counterpropagating circles of black arrows represent the Rashba spin-texture in the InAs quantum well. \textbf{(b)} Fabricated device showing growth sequence of the heterostructure. The Al layer induces a superconducting gap $\Delta^*$, via proximity effect, in the InAs quantum well. \textbf{(c)} Scanning electron micrograph of the array with a scale bar of 1 $\mu$m. \textbf{(d)} Illustrative current-phase relation for a short-ballistic JJ, with high transparency ($\tau$=0.94) and strong SOI, in the absence (black) and presence (red/blue) of an in-plane magnetic field $\boldsymbol{B}_y \parallel \hat{\boldsymbol{y}}$ (red, $B_y>0$; blue, $B_y<0$). The finite magnetic field ($\pm$$B_y$) reduces the critical current by a factor 0.8, $I_c = 0.8I_{c0}$, and adds a cosine term $\pm0.2I_c\cos(\phi)$ to the current-phase relation's Fourier series. The red dots represent the inflection points $(i^*,\phi^*)$ of the current-phase relation. \textbf{(e)} Josephson inductance ($\Phi_0/2\pi I_{c0}$) as function of current ($I_{c0}$), corresponding to the current-phase relation in (d). \textbf{(f)} Resonance curves for the RLC circuit, measured at 500 mK, for different values of the bias current. \textbf{(g)} Current dependence of measured Josephson inductance (at $B=0$). Coloured dots correspond to the spectra in (f). \textbf{(B)} Measurements of inductance and supercurrent anisotropy. \textbf{(a)} Kinetic inductance versus current, for different orientations of in-plane magnetic field of 100 mT. \textbf{(b,c)} Constant (b) and linear (c) coefficients of the polynomial expansion of kinetic inductance L(I) as a function of the angle ($\theta$) between in-plane magnetic field $\boldsymbol{B}_{ip}$ and the supercurrent density directed along $\hat x$. \textbf{(d)} Measured supercurrent magnetochiral anisotropy (coloured lines and symbols) $-2L^\prime_0/(L_0B_{ip})$ versus in-plane magnetic field orientation ($\theta$). The maximum magnetochiral anisotropy, coefficient $\gamma_L\simeq 0.77\times10^6$ T$^-$A$^-$, is extracted from a sinusoidal fit of the data. Fitted supercurrent magnetochiral anisotropy (Grey scale lines) is computed within semiquantitative model (eq. (\ref{CPR})) for different values of the confinement potential $V_{conf}$. The three fitted curves are perfect sinusoidal functions. All measurements are performed at T = 100 mK. Figure is reproduced with permission from ref. \cite{Baumgartner22}.}
\end{figure*}

\subsection{Magnetochiral anisotropy of the inductance}
Unlike fluctuation regime, where nonreciprocity of depairing critical current is tied to the nonlinear resistance, nonreciprocity of sub-Kelvin supercurrent promise fully superconducting/dissipationless nonreciprocal circuit element. Deep in the sub-Kelvin superconducting regime of the phase diagram, i.e., far below the transition temperature where resistance is zero (so DC measurements are not feasible), supercurrent MCA and a corresponding SDE (supercurrent rectification/nonreciprocity) is characterized rather by measuring kinetic (or Josephson) inductance (clearly with AC measurements). By measuring Josephson inductance, nonreciprocal supercurrent can be linked to an asymmetry in the current–phase relation, induced by simultaneous breaking of inversion and time-reversal symmetry such that B is not parallel to I, and the MCA coefficient ($\gamma_L$) for the supercurrent can be directly derived from the equation (\ref{L}). \par

This mechanism can be understood from a semiquantitative model \cite{Baumgartner21,Baumgartner22,Baumgartner22-JPCM} in which Josephson inductance can be derived from the CPR relation $I=I_{c0}f(\varphi)$ (where f is a $2\varphi$-periodic function) and second Josephson equation $\dot{\varphi}=2\pi V/\Phi_0$ (where $\Phi_0=h/(2e)$ is the magnetic flux quantum) as
\begin{equation}
L(I)=\frac{V}{\frac{dI}{dt}}=\frac{V}{\frac{dI}{d\varphi}\dot{\varphi}}=\frac{\Phi_0}{2\pi I_{c0}\frac{df(\varphi)}{d\varphi}} =\frac{\Phi_0}{2\pi}\left[\frac{dI(\varphi)}{d\varphi}\right]^{-1} \;\label{CPR}
\end{equation}
It shows that Josephson inductance is a convenient probe to study CPR symmetry by investigating the effects of space-inversion/time-reversal symmetry breaking on the current–phase relation (CPR). Let's assume a JJ configuration in which electric current is flowing along x-direction, while inversion and time-reversal symmetry is broken by applying out-of-plane electric field $\textbf{E}=E_z\hat z$ and in-plane magnetic field $\boldsymbol{B}_{ip}=B_x\hat x+B_y\hat y$, respectively.

Equation (\ref{CPR}) shows that $L(I)$ is inversely proportional to the derivative of the CPR, therefore, the minimum of Josephson inductance occurs at the inflection-point of the CPR. In the absence of in-plane magnetic field component along y-direction ($B_y=0$), CPR remains symmetric around inflection-point appearing at zero-phase, that is $(i,\varphi)=(0,0)$. As a result, the minimum inductance occurs at zero-current, around which $L(I)$ appears to be symmetric. On the other hand, in the presence of in-plane magnetic field component along y-direction ($B_y\ne0$), CPR become asymmetric around inflection-point $(i^*,\varphi^*)$, mainly associated with the broken Kramers degeneracy between the oppositely polarized spin components of Andreev bound states (ABS) leading to a finite-momentum pairing. As a result, current dependence of the Josephson inductance $L(I)$ also become asymmetric and the minimum of $L(I)$ appears at some finite current $i^*$, corresponding to the shifted inflection point $(i^*,\varphi^*)$ in the CPR. \par

Such a pronounced asymmetry in the skewed CPR and, thus, in the Josephson inductance L(I), signals the supercurrent MCA (as defined in equation (\ref{L})) and hence supercurrent SDE. First, with a given orientation of electric field and polarity of applied current, the shift in inflection point switches along with the sign of $B_y$: $(i^*,\varphi^*)$ for $+B_y$ and $(-i^*,-\varphi^*)$ for $-B_y$, as shown in figure \ref{LI}(Top(d,e)). Second, for a given orientation of $B_y$, the CPR gets more skewed with increasing $B_y$ implying increase in the value of $i^*$ with increasing strength of $B_y$, as shown in figure \ref{LI}(Bottom-a). As a result, as shown in figure \ref{LI}(Top-d), the extremal values of $i^*$ (which are the critical currents $I^+_c$ and $I^-_c$) differ for positive ($\varphi^+_c$) and negative ($\varphi^-_c$) phase difference, signaling the existence of a certain bias-current range in which SDE can be observed for a supercurrent which become different for opposite phase difference polarities. That is, junction allows supercurrent ($I<I^+_c$ (red curve) or $|I|<|I^-_c|$ (blue curve)) along one current direction while it enters in a resistive state ($|I|>|I^-_c|$ (red curve) or $I>I^+_c$ (blue curve)) along the other current-direction.\par

The MCA of the inductance, can be quantified by measuring the constant ($L_0$) and the and the linear ($L^\prime_0$) junction inductance, which appear as the leading terms in the polynomial expansion of L(I) around zero current: $L(I)\approx L_0+L^\prime I+L^{\prime\prime} I^2/2$ with $L^\prime\equiv\partial_I L|_{I=0}$ and $L^{\prime\prime}\equiv\partial^2_I L|_{I=0}$. As shown in figure \ref{LI}(Bottom(b,c)), $L_0$ and $L^\prime_0$ are plotted as functions of the angle $\theta$ between the direction of supercurrent $\boldsymbol{\hat x}$ and the orientation of applied in-plane magnetic field $\boldsymbol{B}_{ip}$. In the Hall-bar geometry of Al/InAs-2DEG/Al junctions with a Ti-Au global top gate, the constant term $L_0$ strongly depends on the gate voltage, reaches its maximum when $B_y=0$, and shows relatively small anisotropy. In contrast, the linear term $L^\prime_0$ shows relatively weak dependence on the gate-voltage, completely vanishes when $B_y=0$ and reaches its maximum when $B_x=0$, and thus shows strongly anisotropic behaviour. As shown in figure \ref{LI}(Bottom-d), MCA coefficient for the inductance $\gamma_L=2L^\prime_0/(L_0B_{ip})$ shows sinusoidal $\theta$-dependence, that is, proportional to $(B\times I)\cdot \hat z=BI\sin\theta$ and agrees with the numerical results obtained from semiquantitative model. In addition, $\gamma_L$ remains nearly independent of the gate-voltage and its maximum extracted from the amplitude of the sine reads  $\gamma_L\simeq 0.77\times10^6$ T$^-$A$^-$. This value of $\gamma_L$, obtained from measurements performed at T = 100 mK, far below the transition temperature ($T_c$), is of the same order as that of $\gamma_S$ calculated for resistive measurements at $T_c$. \par

\section{\label{outlook}Outlook}
SDE is a captivating phenomenon and could be a promising building block of the superconducting dissipationless technologies. Thus far, by characterizing the type/nature of SOI and optimizing/matching the SOI energy with the characteristic energy scale (superconducting gap) of charge carriers \cite{Tokura18}, SDE has been observed in both Rashba SCs and Ising SCs. Recent theoretical studies show that SDE is the strongest (i) when the Cooper pairing momentum lies at the crossover between weak and the strong helical superconducting phase in the vicinity of high critical field, which may be realized via optimizing magnetic field (or intrinsic magnetization), temperature, and SOI \cite{Ilic22} and/or (ii) when the Fermi level lies at the band crossing point of two helical bands, which may be tuned by gating \cite{He22}. From here on, one of the prime goals is to expand the existing platforms and mechanisms for the observation of SDE. For instance, considering the discussion on the optimization of SDE originated from MCS, one of the remaining challenge is to identify suitable superconducting material which may provide the best performance. 
Thus far, in addition to conventional superconducting structures, the SDE has also been predicted and/or observed in unconventional superconducting structures such as twisted few-layer graphene, ferroelectric materials, topological semimetals, and topological insulators. Recent observation of extremely long-range and high-temperature Josephson coupling across a half-metallic ferromagnet \cite{sanchez22-HMJJ} and the prediction of SDE in a JJ with half-metals \cite{SDE-HM22} opens another rout for the search and utilization of promising quantum material class, known as spin-gapless materials \cite{SGS1,SGS2,SGS10,Zengji20}.

In passing, it is interesting to note that the realization of SDE via Rashba SOI and Zeeman exchange interaction in ferromagnetic SCs has a close connection to the realization of QAHE via Rashba SOI and Zeeman exchange interaction in ferromagnetic topological insulators. In the later case, a combined effect of Rashba SOI and Zeeman exchange leads to a spin-splitting in the low-energy bands such that only one of spin sectors display nontrivial band topology with inverted band structure while the other spin sector becomes/remains trivial with normal band structure. As a result, when Fermi level is tuned inside the energy band gap, spin-momentum locked chiral edge state leads to a quantized conductance. In the former case, however, low energy bands in both of the spin sectors play role, mainly due to formation of intra-(Fermi)surface and inter-(conduction)bands spin-singlet Cooper pairing. As a result, when Fermi level is tuned inside the superconducting gap, locking between magnetization orientation and finite-momentum of Cooper pairing leads to finite MCA and nonraciprocity in the supercurrent. Such a fundamental connection between the realization of QAHE and SDE may allow searching suitable topological superconducting materials based on heterostructure of s-wave SCs and QAH insulators \cite{tokura19MTI,Babar19,Nadeem20,Chang22QAH}. In addition, intrinsic iron-based SCs where Rashba SOI-driven band topology and superconductivity coexist \cite{sang22} may also provide promising platform for the realization of SDE in topological superconducting materials \cite{Noah21,Legg22,Takasan22,Dolcini15,Chen18,Kopasov21}.     

However, regarding orientation and the strength of exchange interaction, it is important to remember two differences between the realization of SDE and QAHE. (i) Magnetization orientation needs to be in-plane (at an angel to the polar axis) for SDE while out-of-plane for QAHE. (ii) Nontrivial QAH gap saturates after a critical strength of exchange interaction. On the other hand, strength of SDE decreases after the critical value of exchange interaction $h^*$, yielding crossover between weak and strong helical phase, and vanishes for too high values.\par

On the other hand, considering the reliance of SDE on intrinsic system parameters, search of novel mechanisms may open new rout towards the observation of ideal SDE. In a broader sense, SDE is a manifestation of the interplay between superconductivity and spatial inversion asymmetry. Apart from its realization via MCA induced by time-reversal symmetry breaking, it could also be realized via shift currents induced by nontrivial Berry phase in a time-reversal symmetric systems. Furthermore, for JJs, M. Davydova et al. \cite{Davydova22} recently proposed that finite-momentum Cooper pairing, which elucidates the origin of SDE, can also be achieved without relying on SOI. \par

Similar to the gate-controllability of Fermi level and thus the tunability of SDE strength \cite{He22}, it would be intriguing to understand electric field-effects on the intrinsic properties of a superconducting structure, switching of SDE, and its utilization for dissipationless logic/memory applications. For instance, from the material aspect, SOI, critical current, and pair-breaking are the most important intrinsic properties directly impacting the SDE. Antisymmetric SOI, Rashba and Zeeman SOI, and thus the corresponding spin-splitting can be tuned via electric field. Superconducting pair-breaking shows strong dependence on the strength and the frequency/wavelength of electric field \cite{Chao19}. Similarly, it is shown that a gate tunable critical current in a NbN micro- and nano superconducting bridges \cite{Rocci21} can be enhanced up to 30$\%$. Electric field tunability of superconducting properties has recently been discussed for various ionic-gated superconducting materials, including cuprates, iron-based SCs, and honeycomb structures such as transition-metal dichalcogenides and bilayer SCs \cite{Pengliu22,Piatti21}. For the device prospects, it would be intriguing to replicate magnetic field (or intrinsic magnetization) driven switching of SDE with electric field driven switching via electrical control of magnetization orientation. Electric field driven switching of SDE may also be realized by devising reversible SDE via electric switch of ferroelectricity \cite{Zhai22}. Furthermore, gate-controlled barrier transparency in Rashba semiconductor based JJ (Al/InAs/Al) \cite{mayer20gate} and the gate-controlled asymmetry of highly skewed CPR in topological insulator (BiSbTeSe$_2$) based JJ \cite{kayyalha20gate} demonstrate potential rout of controlling SDE in gate-controlled JJs.\par   

The plausible electric field controllability of SDE and the intertwining between band topology and superconductivity may allow searching new mechanisms/functionalities \cite{phdthesis,Muhd-nano,Muhammad-APR,NCTQFET} of topological quantum materials for steering the engineering of low-power and low-dimensional topological superconducting technologies. We hope this article may provide a route to understand/achieve the optimal performance of SDE and its utilization for superconducting logic/memory device applications.

\begin{acknowledgments}
This research is supported by the Australian Research Council (ARC) Centre of Excellence in Future Low-Energy Electronics Technologies (FLEET Project No. CE170100039) and funded by the Australian Government.
\end{acknowledgments}

\bibliography{apssamp}
\end{document}